\documentclass[12pt,preprint]{aastex}
\newcommand{\HI}{H{\ninerm I}}
\newcommand{\HII}{H{\ninerm II}}

\newcommand{\Klaric}{Klari\'{c}}

\def\Halpha{\hbox{H$\alpha$}}
\def\Hbeta{\hbox{H$\beta$}}
\def\H2{\hbox{H$_2$}}
\def\12CO{\hbox{$^{12}$CO $J = 1\rightarrow 0$}}

\def\deg{\hbox{$^\circ$}}

\def\mJybeam{\thinspace\hbox{mJy\thinspace beam$^{-1}$}}
\def\Jybeam{\thinspace\hbox{Jy\thinspace beam$^{-1}$}}

\def\Jykms{\thinspace\hbox{Jy\thinspace km\thinspace s$^{-1}$}}

\def\K{\thinspace\hbox{K}}

\def\R25{\thinspace\hbox{R$_{25}$}}
\def\Msol{\thinspace\hbox{M$_\odot$}}
\def\Msolpc2{\thinspace\hbox{M$_\odot$\thinspace pc$^{-2}$}}

\def\kms{\thinspace\hbox{km\thinspace s$^{-1}$}}

\def\atc2{\thinspace\hbox{atoms\thinspace cm$^{-2}$}}

\font\ninerm=cmr9               
\shorttitle{CO in NGC 5394/95}
\shortauthors{Kaufman et al.}

\begin{document}
\title{CO Observations of the Interacting Galaxy Pair NGC 5394/95}

\author{Michele Kaufman,\altaffilmark{1}
Kartik Sheth,\altaffilmark{2,3}
Curtis Struck,\altaffilmark{4}
Bruce G. Elmegreen,\altaffilmark{5}\\
Magnus Thomasson,\altaffilmark{6}
Debra Meloy Elmegreen,\altaffilmark{7}
\and
Elias Brinks\altaffilmark{8}
}
\altaffiltext{1}{Department of Physics and Department of Astronomy,
Ohio State University, 174 W. 18th Avenue, Columbus, OH 43210;
e--mail: rallis@mps.ohio--state.edu}
\altaffiltext{2}{Department of Astronomy, University of Maryland, College
Park 20741--2421}
\altaffiltext{3}{present address: Division of Mathematical \& Physical Sciences,
 California Institute of Technology,
MS 105-24, Pasadena, CA 91125; e--mail: kartik@astro.caltech.edu}
\altaffiltext{4}{Department of Physics and Astronomy, Iowa State University,
Ames, IA 50010; e--mail: curt@iastate.edu}
\altaffiltext{5} {IBM Research Division, T.J. Watson Research Center, P.O.
Box 218, Yorktown Heights, NY 10598; e--mail: bge@watson.ibm.com}
\altaffiltext{6}{Onsala Space Observatory, SE--439 92 Onsala, Sweden; 
e--mail: magnus@oso.chalmers.se}
\altaffiltext{7}{Department of Physics \& Astronomy, Vassar College,
 Poughkeepsie, NY 12604;
e--mail: elmegreen@vassar.edu}
\altaffiltext{8}{Depto. de Astronom\'{\i}a, Universidad de Guanajuato,
Apdo. Postal 144, Guanajuato, Gto. 36000, M\'exico; e--mail:
ebrinks@astro.ugto.mx}

\begin{abstract}
BIMA \12CO\ observations are presented of the 
spiral galaxies NGC 5394 and NGC 5395 that have undergone a recent,
grazing encounter. In NGC 5394, approximately 80\% of the CO emission detected
by BIMA is concentrated in the central 800 pc (FWHM) starburst region, and the
rest is from a portion of the inner disk south and west of the central
starburst. In an encounter simulation that reproduces some of the main features
of this galaxy pair, a considerable amount of gas in NGC 5394 falls into the
central region early in the collision. The observed total gas distribution 
in the disk of NGC 5394 is lopsided, with more \HI, CO, and \Halpha\ emission  
coming from the western or southwestern side.
The innermost western arm of NGC 5394 is seen in CO and \Halpha\ emission, 
but the eastern inner--disk arm, which is very bright in the optical 
continuum, is not detected in CO or \Halpha\ emission. 

The NGC 5394 starburst region is similar in radio continuum luminosity and 
size to the M82 starburst and has a CO luminosity $\sim 4$ 
times greater.
A CO position--velocity diagram of the NGC 5394 nucleus reveals 
two separate velocity features very close to the center.
This may indicate a nuclear ring or the ``twin peaks'' of an ILR or
some depletion of \12CO\ at the nucleus.
From a comparison of the radio continuum,
\Halpha, 60 \micron, and CO luminosities, we estimate that the average 
extinction
$A_{\rm v}$ of the starburst nucleus is 3 -- 4 mag, the star formation
rate is $\sim 6$ \Msol\ per year, and
the conversion factor $N({\rm H_2})/I_{\rm CO}$ in the starburst
is a factor of 3 -- 4 below the standard value.  Comparison of NGC 5394
with two other systems previously studied
suggests that in prograde grazing encounters a central starburst may not
develop until near the end of the ocular phase.

Very little of the CO emission from NGC 5395 found in previous 
single--dish observations 
is detected in the BIMA data; thus molecular gas in NGC 5395 does not appear 
to be strongly concentrated in compressed ridges.

\end{abstract}


\section{Introduction}
\begin{figure}
\epsscale{0.95}
\plotone{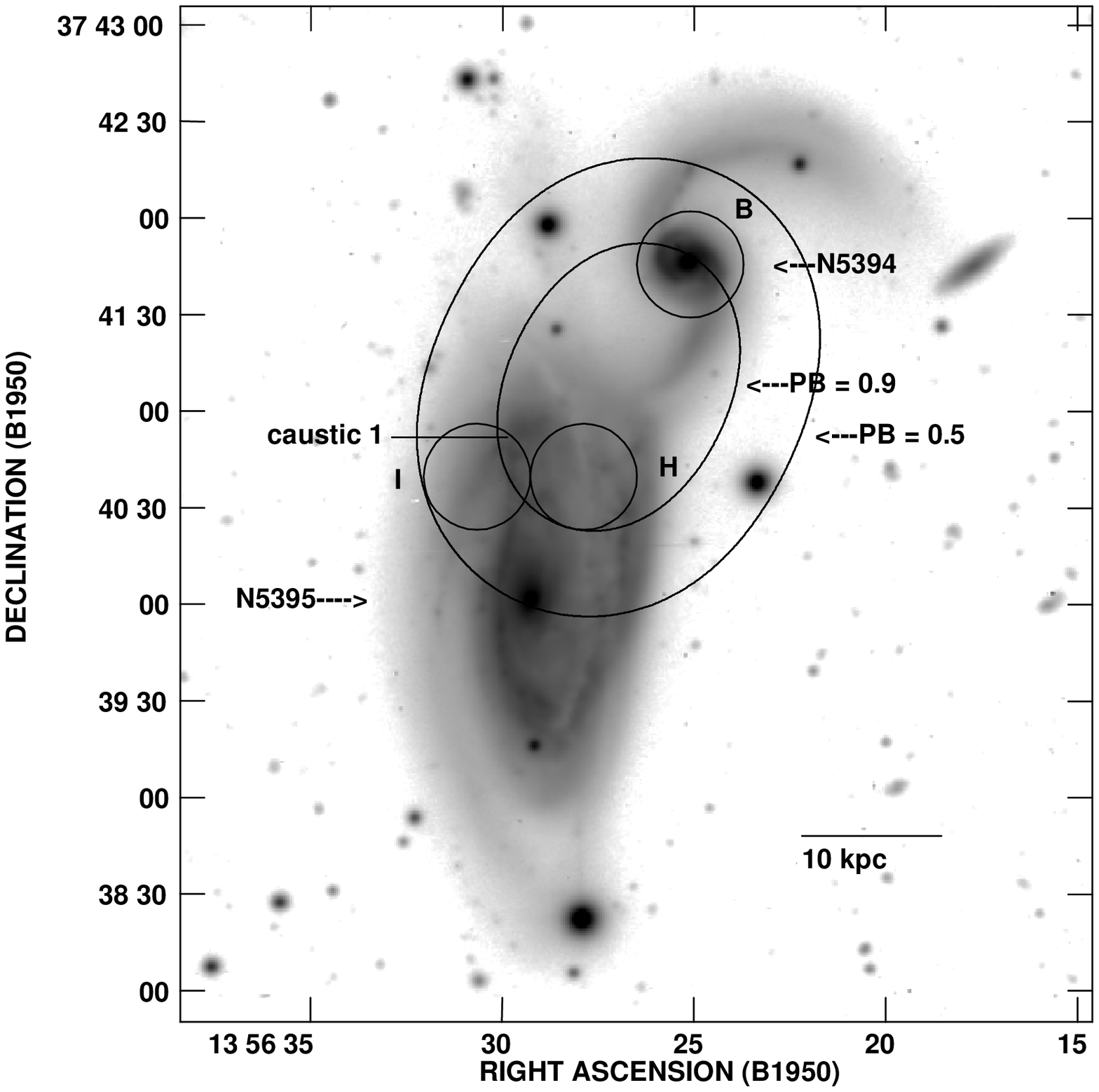}
\caption{The $r$--band image of NGC 5394/95 (on a logarithmic intensity scale) 
from the Chengalur--Nordgren galaxy survey. 
The circles labelled B, H, and I
denote the Onsala HPBW at the positions where CO emission was 
clearly detected by Onsala Space Observatory.
Position B contains the bright inner--disk arms and starburst nucleus of
NGC 5394. The field observed in
\12CO\ in the BIMA observations presented here is indicated by 
the two oval contours marked ``PB,'' which represent the 50\% and the 
90\% sensitivity contours of the combined primary--beams after the two 
pointings were mosaiced. In NGC 5394 all of the CO emission clearly detected 
in the BIMA 
data is from the central and southwestern portion of Position B. The stellar
feature in NGC 5395 labelled ``caustic 1'' was interpreted in Paper I as a
caustic produced by temporary convergence of orbits. 
\label{fig:cornell}
}
\end{figure}

The interacting spiral galaxies NGC 5394/95 (Arp 84) at a distance of 47 Mpc
 have recently undergone a 
nearly grazing encounter. Figure~\ref{fig:cornell} displays an $r$--band 
image of the system from the Chengalur--Nordgren galaxy survey  
\citep{NOR97}.
From a comparison of the observed tidal features of the system with galaxy 
encounter simulations, \citet{KAU99} (hereafter 
Paper I) concluded
that NGC 5394, the smaller member of the pair, suffered a close
prograde collision with NGC 5395 and is in an immediate post--ocular 
(post eye-shaped) phase. NGC 5394
has a central starburst, 3$\farcs$5 (= 800 pc) in diameter, very bright 
inner-disk spiral arms ($R = 4'' - 13''$ = 0.9 -- 3.0 kpc from the nucleus), 
and two long, open tidal arms, with the tidal tail extending to  
$R = 80''$. Its disk is
intrinsically oval. Some properties of NGC 5394 are listed in Table 1. 
The dominant spiral arm of its companion, NGC 5395,  forms a large ring or
pseudo-ring of H$\alpha$,
radio continuum, and \HI\ emission. The eastern half of the ring is the 
brighter side in radio continuum and H$\alpha$, whereas the western half is 
the brighter side in \HI.

 NGC 5394/95 is one of three galaxy pairs that
we have studied previously  with radio, optical, and single--dish CO
observations and numerical simulations to see how structures resulting 
from prograde, in--plane, grazing encounters develop. The other two systems are
IC 2163/NGC 2207 \citep{ELM95a,ELM95b,ELM00,ELM01} and
NGC 2535/36 \citep{KAU97}. In the models, the 
response of the 
galaxy disk to the $\cos 2 \theta$ tidal potential of the prograde encounter 
produces an eye--shaped (``ocular'') oval like that seen 
in IC 2163 and NGC 2535. This can subsequently evolve into 
inner--disk arms, similar to those visible in NGC 5394. From a 
comparison of the observed morphologies with the models, Paper I concluded 
that the galaxies IC 2163, NGC 2535, and NGC 5394 form an ``age'' sequence with
IC 2163 the least evolved interaction and NGC 5394, the most evolved one.
IC 2163 has star formation typical of normal spiral disks
\citep{ELM01}, NGC 2535 has somewhat enhanced star 
formation \citep{BER93} with the brightest \Halpha\
emission at the two ends of the eye--shaped oval 
\citep{AMR89},
and NGC 5394 has a central starburst.

We present \12CO\ observations of NGC 5394/95 taken with the 
Berkeley--Illinois--Maryland 
Association (BIMA)\footnote{The BIMA array is partially funded by grant
AST--9981289
from the National Science Foundation} array to examine the distribution
of molecular gas, particularly in 
the inner--disk arms and starburst nucleus of NGC 5394 and in
the ring/pseudo-ring of NGC 5395.

 The inner--disk 
arms of NGC 5394 present a puzzle for models of star formation. 
These bright arms are mostly free of star-formation knots 
\citep{ARP69,WRA88}, and only one of the 
three arms is clearly detected in H$\alpha$ (Paper I). This is unusual, since
high surface brightness in a spiral arm is almost always the result of
ongoing star formation. \citet{WRA88}  described these arms as 
extremely bright spiral
arcs that are ``mostly smooth and green.'' Despite the presence of \HI\
gas, two of these arms show no evidence of star formation in progress. Paper I
offered the following interpretation.
 If molecular gas can be ignored, the instability
parameter $Q_{\rm gas} > 10$, which is too high for significant 
star formation; \citet{KEN89} and \citet{MAR01}  found that 
$Q_{\rm gas} \leq 1.4$
for star--forming regions in late--type spirals.  
A value for $Q_{\rm gas}$ based simply on
\HI\ may be too high if a significant amount of molecular gas, which
can lower the value of $Q_{\rm gas}$, is present.
 Thus we needed to measure the 
 molecular surface density at the inner--disk arms of NGC 5394.
Single--dish observations with the NRAO 12--m telescope at a resolution of
$55''$ \citep{ZHU99} and with Onsala Space Observatory 
at a resolution of $33''$ (Paper I) detected
significant CO emission from NGC 5394. In particular, the Onsala observations
found a lot of \12CO\ emission (129 Jy \kms) in the field containing 
the starburst nucleus and the inner--disk spiral arms but did not have
sufficient spatial resolution to distinguish between emission from
the nuclear region and that from the inner--disk arms (Paper I).
The 5 times higher resolution ($\sim 6 \farcs 5$) of the BIMA \12CO\ 
observations 
presented here allows us to make this distinction.

Optical spectroscopy by \citet{ARP69}  and
\citet{LEE89} and near--infrared spectroscopy by
\citet{VAN98} revealed a young starburst in the 
center in NGC 5394. The position of NGC 5394 in the near--infrared
line--ratio diagnostic diagrams
indicates a starburst of
duration a few Myr, observed several Myr after the peak star--formation 
episode. The starburst in NGC 5394 has nearly the same radio continuum
luminosity and linear size as the central starburst in
M82 (Paper I; \citet{CON90}).  Unlike M82, 
NGC 5394 is viewed nearly face--on. The \Halpha\ line
profile at the nucleus of NGC 5394 has a 100 \kms\ velocity asymmetry
(an extended blue wing), which may represent a wind/outflow of ionized gas
perpendicular to the disk (Paper I). 
In the M82 starburst, an outflow speed of several hundred \kms\ is seen in the 
molecular gas
\citep{SHE95}  as well as in the
ionized gas.
Our BIMA observations of NGC 5394, particularly the velocity data,
provide information about the structure of the starburst region in NGC 5394 and 
the relative spatial distributions of \12CO\ and 
H$\alpha$ emission.

For the ``standard'' value of the Milky Way
conversion factor $X = N(\H2)/I_{\rm CO}$, we adopt  
$2.3 \pm 0.3 \times 10^{20}$  \H2\ cm$^{-2}$ (K \kms)$^{-1}$  
from \citet{STR88} since this value has been widely 
used. More recent determinations for the Milky Way give somewhat
lower values of $1.9 \pm 0.2 \times 10^{20}$ 
\citep{STR96}, and $1.8 \pm 0.3 \times 10^{20}$ 
\citep{DAM01}. We shall estimate the value of $X$ for
the starburst region of NGC 5394 by comparing its radio continuum, \Halpha,
60 \micron, and CO luminosities. For
luminous infrared galaxies involved in mergers, measurements of the dynamical
mass \citep{ALO01,SHI94} and estimates of the molecular mass based
on millimeter--wave thermal emission of dust 
\citep{BRA98} indicate that the standard value of
$X$ overestimates the molecular mass by factors ranging from 3 to greater
than 10 in these powerful starbursts. The starburst in NGC 5394 is more 
modest, comparable to that in M82. \citet{WIL92} and
\citet{SMI91} found that the value of $X$ in the 
M82 starburst is a factor of 3 below the standard value.

The northern half of NGC 5395 was included as well in our BIMA observations.\\
\12CO\ emission from NGC 5395 was detected in the Onsala observations
(Paper I) and in the NRAO 12--m observations by \citet{ZHU99}.
 In other galaxies, nonthermal radio continuum emission
and \12CO\ emission are usually correlated on scales of $\sim 2$ kpc
\citep{ADL91}.
Since the eastern half of the ring/pseudo--ring in NGC 5395 is the brighter 
side in radio continuum emission but the fainter side in \HI,
we wanted to find out if the eastern half of the ring is the brighter
side in CO emission. We expected
to see a strong shock front in the molecular gas along the radio continuum
ridge. We also looked for any CO emission associated
with an unusual ``shell--like'' stellar structure on the northeastern side
(marked as ``caustic 1'' in Fig.~\ref{fig:cornell}), which was 
interpreted in Paper I as a caustic produced by temporary 
convergence of orbits.  

In addition to the BIMA CO data, our analysis uses \HI, Fabry--Perot
\Halpha,
radio continuum, single--dish CO, and broad--band near-infrared and optical
observations from Paper I and the VLA radio continuum image from the FIRST
survey \citep{BEC95}.

With the adopted distance of 47 Mpc for the galaxy pair, $1''$ corresponds
to 230 pc.  The velocities in
this paper are heliocentric and use the optical definition for the
nonrelativistic Doppler shift.\footnote{Where necessary, the relation
$v_{\rm helio,opt}$ = $v_{\rm lsr,radio}$ + 28 \kms\ was used to convert 
velocities to the heliocentric, optical system. Recall that the optical 
definition is 
$v_{\rm opt}$ = $c(\lambda - \lambda_0)/\lambda_0$, where $\lambda_0$ is the
rest wavelength.} 

Section 2 describes the observations and the data reduction. 
Section 3 describes the optical and near--infrared morphology of NGC 5394.
Sections 4, 5, 6,
and 7 present our CO results on NGC 5394, with the general CO distribution in 
Section 4, the eastern inner--disk arm in Section 5, the central starburst in 
Section 6, and a comparison of CO, \Halpha, and \HI\ velocity fields in 
Section 7. Section 8 discusses the implications for development of a 
starburst in a nonmerging encounter of this type. Section 9 discusses  NGC 5395.
Section 10 contains our conclusions.

\section{Observations and Data Reduction}

The BIMA \12CO\ observations were made on 1999 June 4 with 10 telescopes in C 
configuration. We used two pointings to cover NGC 5394 and
the northern half of NGC 5395. The duration of the run was 10.7 hours.
Table 2 gives a summary of the observations.  The phase calibrator was 
the quasar 1310+323, the bandpass calibrator was 3C 273, and the absolute 
flux standard was Mars. The observations have a
velocity resolution of 4.06 \kms\ and were centered on 3497 \kms\ with
a total bandwidth of 908 \kms.
 
\begin{figure}
\epsscale{0.86}
\plotone{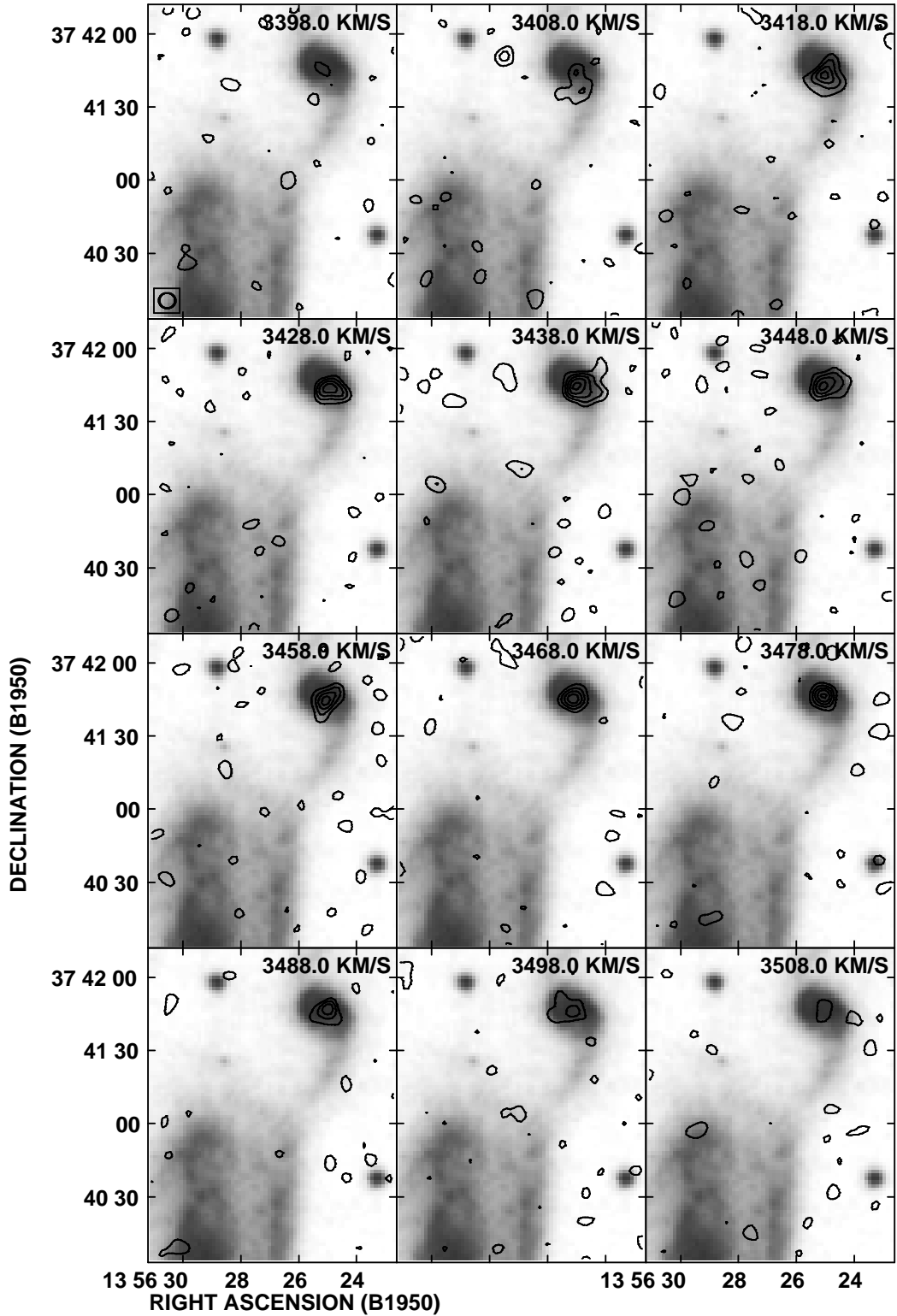}
\caption{Channels maps from the BIMA CO cube overlaid as contours
 on a gray--scale display of the Digitized Sky Survey image for the range 
of velocities appropriate to NGC 5394. The contour levels are 
2, 4, 6, 8, and 10 times the rms noise of 0.18 K.
\label{fig:cube} 
}
\end{figure}

The data reduction was carried out using standard reduction algorithms from 
the MIRIAD package \citep{SAU95}. The data were Hanning smoothed to
10 \kms\ channels. The two pointings were mosaiced to form a single cube 
with pixel 
size = $1 \farcs 5$, synthesized beam = $6\farcs 6  \times 6\farcs 3$, and
dimensions $198 \times 211 \times 60$ pixels 
(R.A. $\times$ DEC $\times$ velocity),
covering the velocity range 3208 to 3808 \kms. 
The dirty channel maps were then deconvolved by using
the Steer--Dewdney--Ito CLEAN algorithm \citep{STE84}.
  The rms noise in the cleaned
10 \kms\ channel maps is 80 \mJybeam, only slightly higher than the theoretical
noise of 76 \mJybeam. In Fig.~\ref{fig:cornell}, the 50\% and 90\%  
sensitivity contours of the combined primary beams after mosaicking the
data are overlaid on the $r$-band image in grayscale.
Figure~\ref{fig:cube} displays channel maps from the BIMA CO cube overlaid
on a gray--scale display of the Digitized Sky Survey image\footnote{The
POSS Digitized Sky Survey was produced at the Space Telescope Science
Institute under grant NAG W-2166.} for the range of velocities appropriate
to NGC 5394. Moment maps of the total emission
and of the intensity--weighted velocity field were made from
this cube.

Areas of genuine CO emission were selected by clipping at $2 \times$ the
rms noise in a cube convolved to $12''$ resolution and requiring that a 
feature be continuous in velocity in at least two adjacent 
channels. In NGC 5394, the only \12CO\ emission seen in the BIMA 
observations is
from the central and southwestern half of the Onsala position B
marked in Fig.~\ref{fig:cornell}, 
that is, only from the central part of the galaxy. The CO emission from 
NGC 5394 is present over the velocity range 3388 to 3528 \kms, and the 
peak brightness temperature $T_{\rm b}$ in the channel maps is 2.05 K
(after correction for primary beam attentuation). BIMA finds very little
CO emission from NGC 5395; a small region 
inside the arc formed by the
stellar caustic has a peak $T_{\rm b}$ of 0.87 K (= $4.8 \times$ the rms
noise) at $v = 3308$ \kms.

\section{Optical and Infrared Morphology of NGC 5394}

NGC 5394 has $M_{\rm B} = -19.7$. On the basis of the \HI\ velocity field,
the observed value of $v_{\rm max} \sin i$, and 
the  $I$--band luminosity, Paper I concluded that NGC 5394 is viewed nearly 
face--on with its major axis of projection
into the sky--plane at a position angle of 0\deg. 
The stellar disk of NGC 5394 is intrinsically oval as a result of the
encounter. In the sky--plane, the position angle of the photometric major 
axis varies continuously from 60\deg\ to --6\deg, and the ellipticity 
increases from 0.21 to 0.60 (Paper I) as the galactocentric distance  
increases from $13''$ (the start of the tidal arms) to 
$51''$ (the isophotal major radius $R_{25}$ measured by \citet{DEV91}).
 Ignoring the small difference at $R_{25}$ between the position
angles of the photometric major axis and the projection line of nodes,
the area of the face--on disk at the 25th magnitude isophote =  
$\pi a_0 b_0$ = $\pi (b \sec i/a ) R_{25}^2$, where $a, b$ are the semi--major
and semi--minor axes in the sky--plane and $a_0, b_0$ are the face--on 
values. For comparisons with other galaxies, as in Section 8 below, we 
take the effective value of $R_{25}$ as $(R_{25})_{\rm eff}$ = 
$(b \sec i/a)^{1/2} R_{25}$. With inclination $i$ = 15\deg\ and $b/a$ = 0.4,
this gives $(R_{25})_{\rm eff} = 33''$ = 7.6 kpc.

\begin{figure}
\epsscale{1.18}
\plottwo{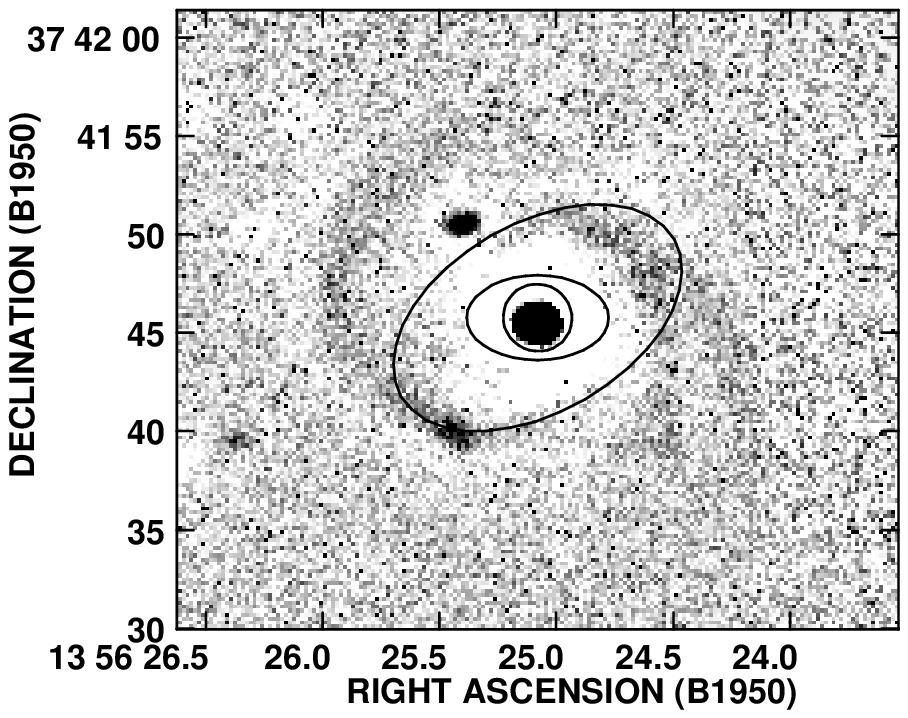}{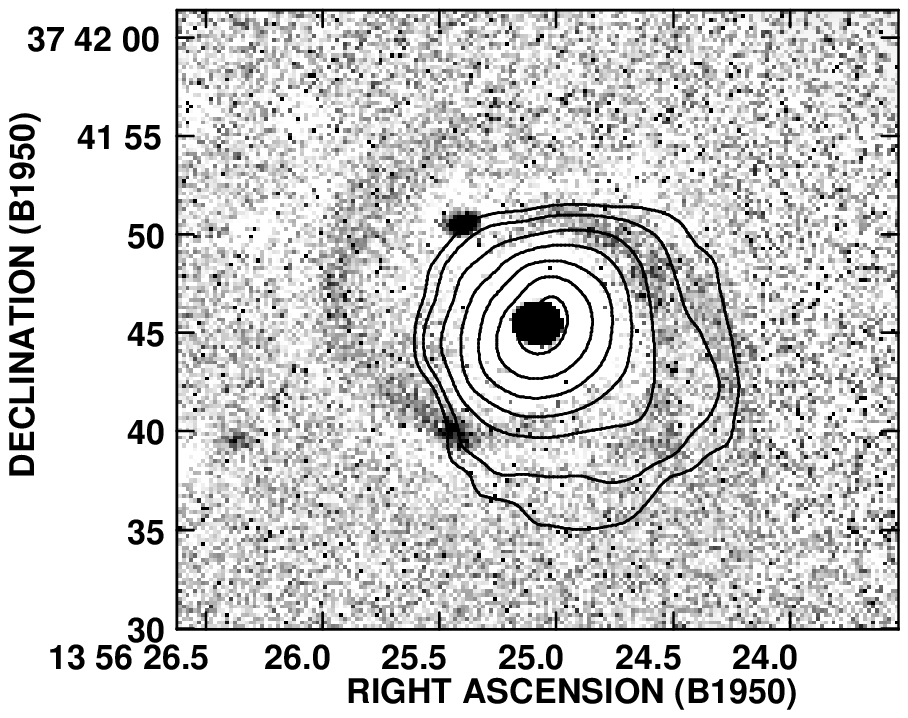}
\caption{Top: Sketch of features visible in optical and near--infrared images
of the center of NGC 5394. An unsharp--masked $J$--band image, which shows
the inner--disk arms, is displayed in gray--scale. The western inner--disk arm 
appears to bifurcate at a position angle of --50\deg\ into what we call the 
two western inner--disk arms. The larger of the two oval contours
represents the primary bar--like feature with a semi--major axis of 
$8''$; the smaller oval contour, a possible secondary bar with a
semi--major axis of  $3\farcs 5$.
The central circle indicates the extent of the starburst.
Bottom: Contours of BIMA CO integrated--intensity, after correction for
primary--beam attentuation, overlaid on the 
unsharp--masked $J$--band image in gray--scale. 
 The $I_{\rm CO}$ contours are at 5, 10, 20, 30, 40, 50, 60 \Jybeam \kms, 
where 10 \Jybeam \kms\ = 22 K \kms, and the rms noise is 2.8 \Jybeam \kms.
At this location, the correction factor for primary--beam attenuation 
in the mosaiced BIMA data is 1.08. 
No CO emission is 
detected from the eastern inner--disk arm, which is bright in the optical
continuum. 
\label{fig:J-band}
}
\end{figure}

NGC 5394 has two long, open tidal arms (see Fig.~\ref{fig:cornell}) 
with high arm--interarm contrast
(Paper I). The very bright inner--disk spiral arms are disjoint from the long
outer tidal arms, as the latter depart from the side of the inner--disk
arms at a large angle. The top panel of Figure~\ref{fig:J-band} is a 
sketch of features visible in the inner disk and center of NGC 5394. 
The eastern inner--disk arm spirals outwards for 210\deg\ in a clockwise
direction, starting at $R = 4''$ at a position angle of 200\deg\ and 
ending at $R = 13''$.
 The western inner--disk arm spirals outward from
$R = 4''$ at a position angle of 30\deg\ and appears to bifurcate at
a position angle of --50\deg\ into what we call the two western inner--disk
arms. The outer of the two western inner--disk arms is symmetric to the
eastern inner--disk arm, whereas the innermost western arm seems to be
a separate structure.
There is no evidence of on--going star formation in either the eastern
or the outer of the two western inner--disk arms. \Halpha\
emission is detected from the innermost western arm (Paper I).
   
There is a barlike structure (see Fig.~\ref{fig:J-band}, top panel)
 with a major axis of $16''$ (= 3.7 kpc) at a position angle of 120\deg\
(Paper I; \citet{BUS92}). We shall refer to this
feature as the primary bar; with an axis ratio of $\sim 1.7$, it is a
fat bar. The inner--disk arms
appear tangent to part of the primary bar but do not depart quite
from the ends of the bar. In the optical images, such as the
Chengalur--Nordgren $r$--band image \citep{NOR97} in Fig.~\ref{fig:cornell} 
which has $2 \farcs 3$ resolution,
one does not see straight dust lanes along the primary bar.
Straight dust lanes are the signature of a bar that
extends between an inner Lindblad resonance and corotation 
\citep{ATH92a,ATH92b}. Measured 
widths of dust lanes in galaxies with large bars are, typically, 100 -- 250 pc 
from \citep{BEN82} 
on NGC 7479 and our measurements on NGC 1300 and NGC 5236
(from images in \citet{ELM96} and 
\citet{SAN88},
 respectively). At the distance of NGC 5394, this corresponds to
$0 \farcs 5$ -- $1 \farcs 0$, so, if present in NGC 5394, the optical 
resolution may not be good enough to see them.
Interior to the primary bar, the near--infrared $J$
and $K$--band images by \citet{BUS92} 
 reveal an elongated 
feature with a major axis of $\sim 7''$ and a position angle of 90\deg,
thus trailing the bar by 30\deg\ in azimuth. This smaller oval feature
is not evident in the
$J$--band image ($1''$ resolution) in Paper I or the 2-MASS $J, H, K$
images ($3 \farcs 3$ resolution) or in the Chengalur--Nordgren $r$--band
image. In Section 7 below, we consider the 
kinematic evidence for this possible secondary bar.  

From the twist of the \HI\ velocity contours at the outer tidal arm, 
Paper I concluded that the outer tidal arms are outside of corotation. 
If the outer arms are driving a density--wave in the inner arms
(it is not clear that this is occurring), then
the radius where the inner--disk arms end and 
the outer tidal arms begin (1.6 times the semi--major axis of the primary 
bar) may represent either corotation or the outer Lindblad resonance for
the inner--disk spiral arms.

\section{General CO Distribution in NGC 5394}

The Onsala observations (Paper I) measured a total 
integrated \12CO\ flux $S_{\rm CO}$ of 168 Jy \kms\
from a five--position survey of NGC 5394 with a $33''$ (HPBW) beam at
$33''$ spacing; of this, 129 Jy \kms\ is from the pointing 
(position B in Fig.~\ref{fig:cornell}) centered on the starburst 
nucleus and the rest is from a weak detection ($3 \times$ rms noise) 
in a pointing centered $33''$ south of the nucleus,
which includes the southern tidal arm. With the NRAO 12--m telescope,
\citet{ZHU99} measured a CO integrated flux of
154 Jy \kms\ in a $55''$ (HPBW) beam; this is consistent with the
Onsala value. With the adopted standard value of the conversion factor $X$, 
the Onsala  $S_{\rm CO}$ 
corresponds to an $M(\rm H_2)$ of $3.4 \times 10^9$ \Msol. 
Paper I found $M$(\HI) = $7.3 \times 10^8$ \Msol. After adding 
a 36\% by mass contribution from helium to the atomic plus molecular
hydrogen mass, this gives a total mass in gas
of $5.5 \times 10^9$ \Msol. However, as mentioned in the Introduction, use
of the standard value of $X$ in a starburst region is questionable.   
We shall estimate $X$ for the starburst region in Section 6 and discuss 
the $M({\rm H_2})/M({\rm HI})$ ratio for the galaxy in Section 8 below.

\begin{figure}
\epsscale{1.25}
\plottwo{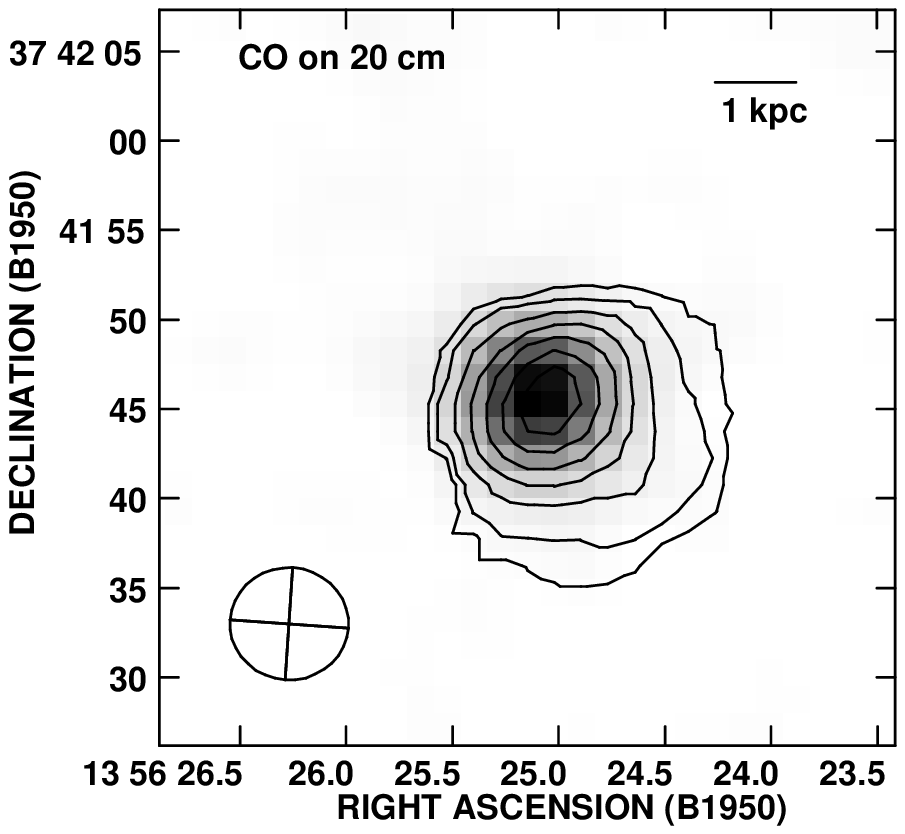}{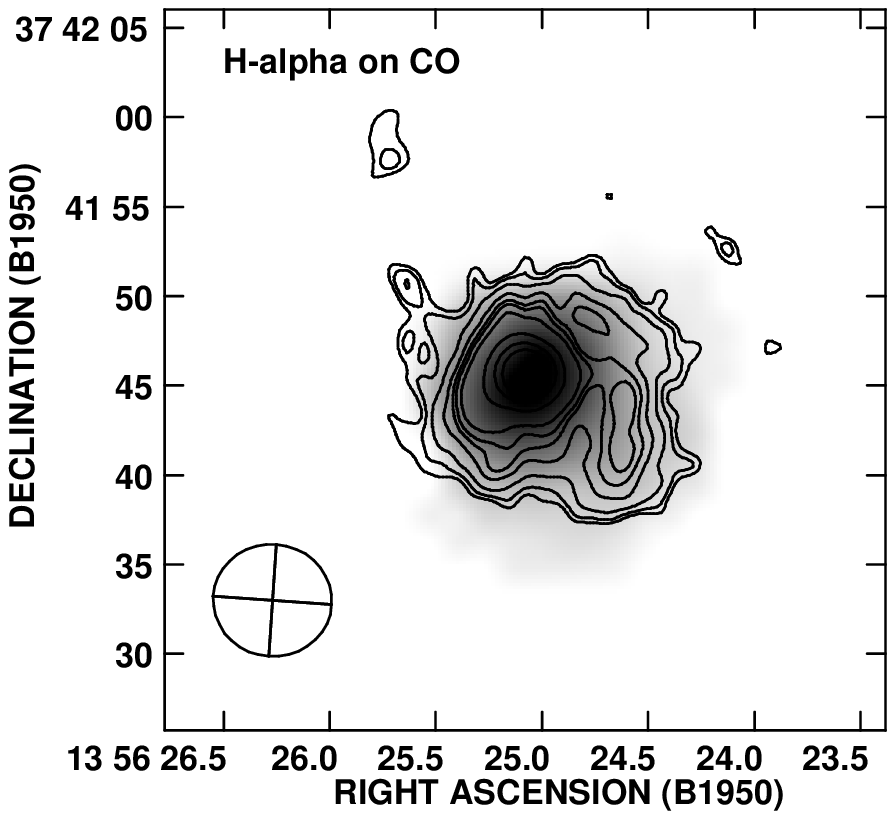}
\caption{Top: Contours of BIMA \12CO\ integrated--intensity 
overlaid on the radio continuum $\lambda = 20$ cm
image of the center of NGC 5394 from the FIRST survey in gray--scale.
Bottom: Contours of H$\alpha$ emission overlaid  on the BIMA CO 
integrated--intensity image of the center of NGC 5394 in gray--scale.
In the inner disk, CO and \Halpha\ emission have
a similar lopsided spatial distribution with more emission from the 
southwestern side. In both panels, the beam symbol represents the CO resolution.
\label{fig:I(CO)}
}
\end{figure}

In the bottom panel of Figure~\ref{fig:J-band}, contours
of the BIMA \12CO\ 
integrated--intensity $I_{\rm CO}$, after correction for primary--beam
attenuation, are overlaid on an unsharp--masked
$J$--band image ($1''$ resolution, from Paper I) in gray--scale to show the 
location of the
CO emission relative to the eastern inner--disk arm and the two western
inner--disk arms. The bottom panel of Fig.~\ref{fig:I(CO)}
displays contours of H$\alpha$ emission ($2 \farcs 4$ resolution, from
Paper I) overlaid on the $I_{\rm CO}$ image in gray--scale.
In the nucleus there is a marginal displacement between the
surface brightness maximum in CO and the optical/radio continuum surface
brightness maximum (see Section 6 below).
Aside from this, the \12CO\ emission and H$\alpha$ emission have a similar
lopsided distribution of extended emission towards the southwest.
In particular, no CO
or H$\alpha$ emission is detected from the eastern inner--disk
arm, which is bright in the optical continuum. 
Star formation in the disk appears to follow 
the distribution of molecular gas.  

The total integrated CO flux 
detected with BIMA is $S_{\rm CO}$ = 117 \Jykms, with an rms noise of
$\pm 8$ \Jykms\ and an uncertainty in the flux calibration of 
10\% - 15\%. For the aperture
containing the starburst nucleus and the inner disk arms (position B), 
the single--dish
Onsala observations find $S_{\rm CO}$ = 129 \Jykms, with an rms noise of
$\pm 7$ \Jykms. Thus,  
within the uncertainties (rms noise, calibration), the BIMA map detects 
all the CO emission
seen by Onsala at position B, so the Onsala observations support the conclusion
from the BIMA map that any CO emission at the eastern inner--disk arm is 
below the rms noise in either data set.

In the top panel of Fig.~\ref{fig:I(CO)}, $I_{\rm CO}$ contours 
are overlaid on the radio continuum $\lambda = 20$ cm image from
the FIRST survey \citep{BEC95} in 
gray--scale. The spatial resolution of the BIMA image 
($6 \farcs 6 \times 6 \farcs 3$, HPBW) is
a little lower than that of the radio image ($5 \farcs 4$ HPBW). The radio
continuum emission is strongly peaked on the starburst nucleus and
symmetrically distributed about it. 

\begin{figure}
\epsscale{0.6}
\plotone{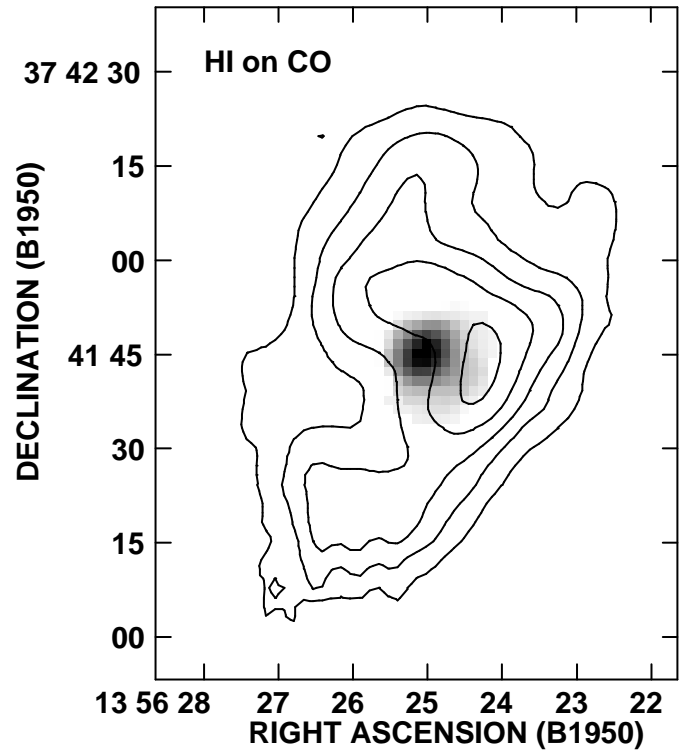}
\caption{Contours of the $N$(HI) image (resolution $18''$) from Paper I overlaid
on the BIMA $I_{\rm CO}$ image of NGC 5394 in gray--scale. This displays the
larger scale \HI\ structure of the galaxy whereas Figs. 3, 4, and 6 are
enlargements of the central region. 
The $N$(HI) contour
levels are at 2.0, 4.3, 4.8, 6.1, and $7.5 \times 10^{20}$ \atc2. 
\label{fig:HI}
}
\end{figure}

The \HI\ observations of Paper I are more sensitive to low column density
gas than the CO observations presented here. We detected \HI\ emission
from the entire inner disk, the southern tidal arm, and the inner third of
the northern tidal arm (see Figure~\ref{fig:HI}). The line--of--sight \HI\ 
column density $N$(HI) is greater on the western side than on the eastern
side of the disk. The distributions of
the ionized gas, atomic gas, 
and molecular gas in the disk of NGC 5394 all favor the western or
southwestern side.
 Unless driven by external forces, such a lopsided distribution
of gas should not last more than a few shear times.

\begin{figure}
\epsscale{1.25}
\plottwo{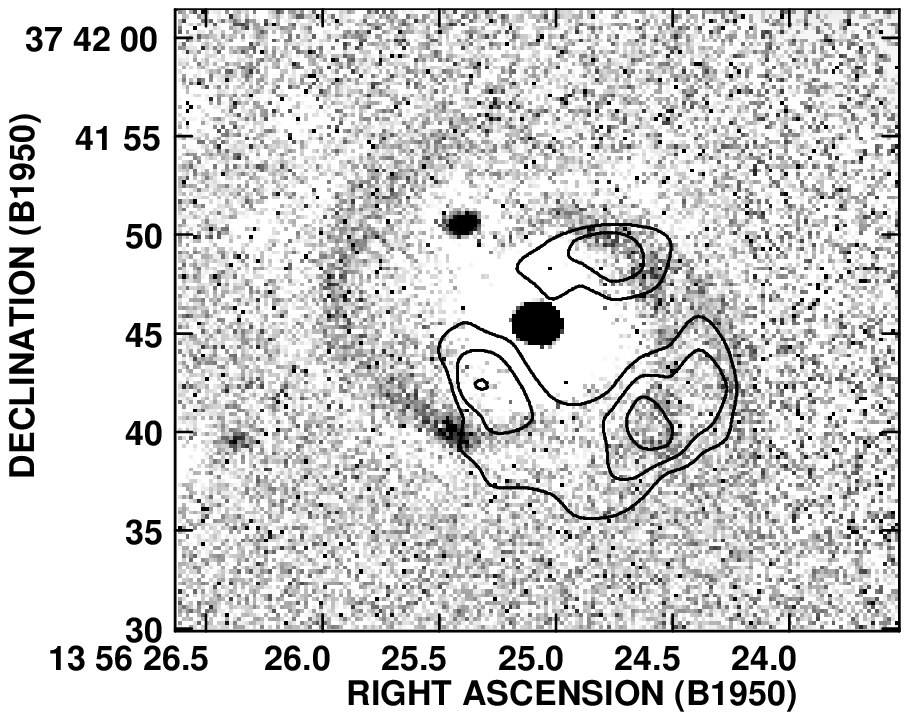}{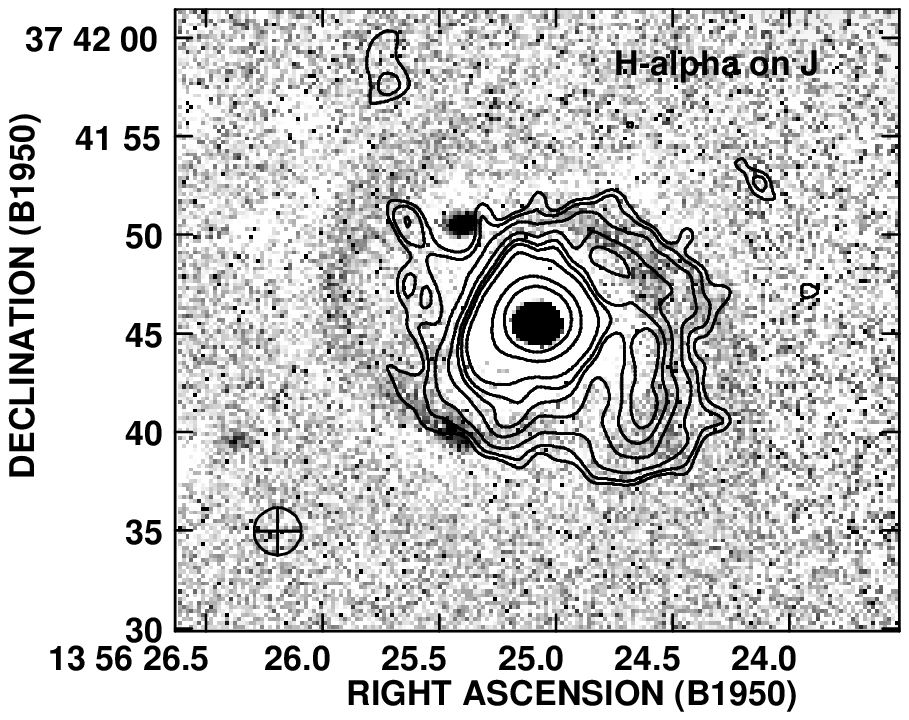}
\caption{Top: Contours of the residual $I_{\rm CO}$ from NGC 5394 after 
subtracting the emission from the central starburst (assumed axisymmetric). 
These are overlaid on the unsharp--masked $J$--band image in gray--scale to
indicate the correspondence between the residual CO--emission and part of
the innermost western arm. The $I_{\rm CO}$ contours are at 6, 9, 12
\Jybeam \kms\ and the rms noise is 2.8 \Jybeam \kms.
Bottom: Contours of \Halpha\ emission overlaid on the unsharp--masked
$J$--band image in gray--scale. The beam symbol represents the \Halpha\
resolution.
\label{fig:residual}
}
\end{figure}

\begin{figure}
\epsscale{1.0}
\plottwo{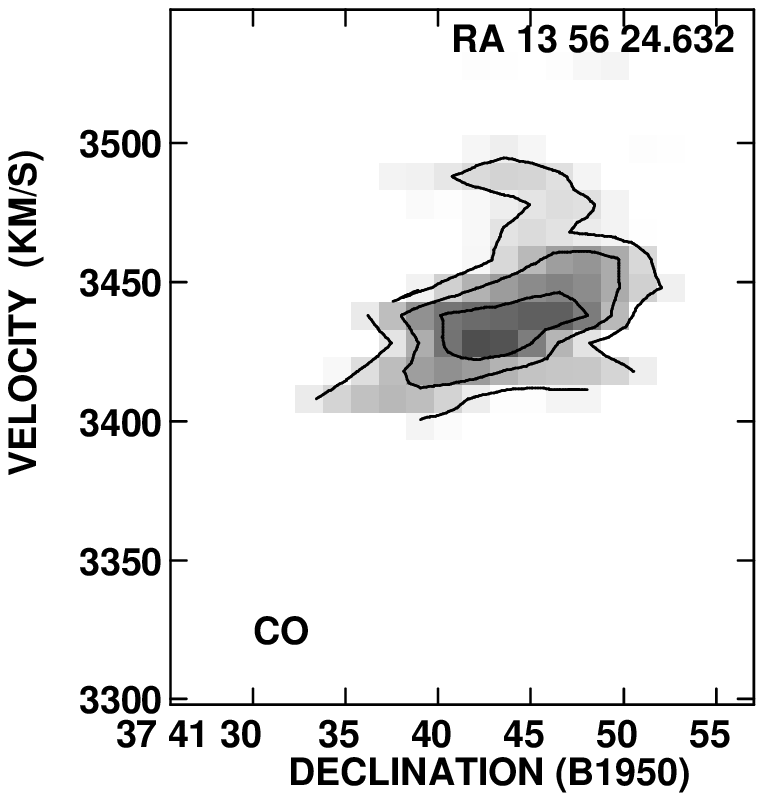}{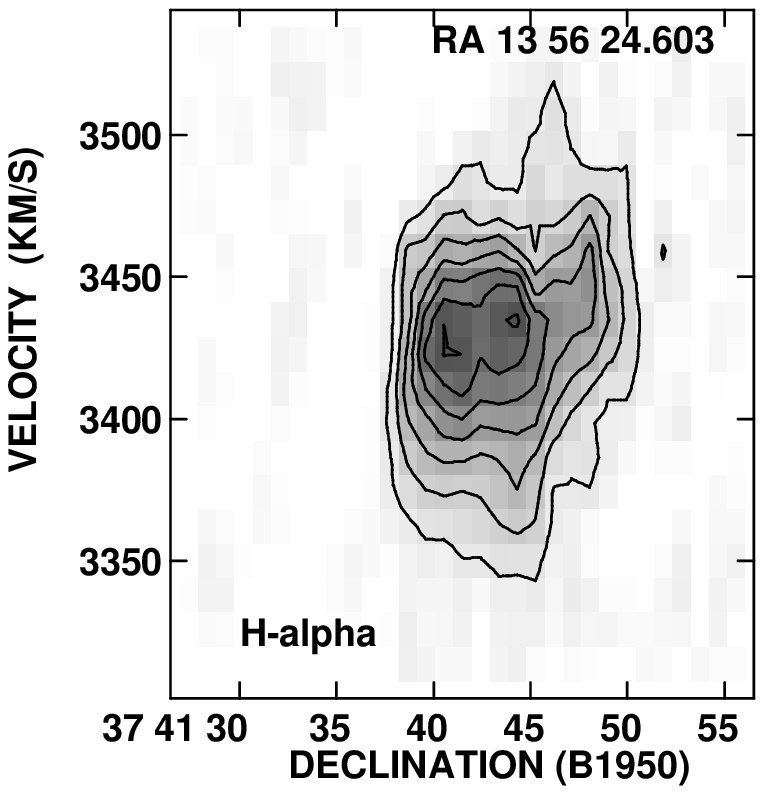}
\caption{Position--velocity diagram in CO (left panel) and
\Halpha\ (right panel)
of the emission from the innermost western arm of NGC 5394. 
The cut is along the declination axis, $6''$ west of the nucleus.
The contour levels in CO are at 2, 4, and 6 times the rms noise of 
80 \mJybeam. The contour levels in \Halpha\ are at 4, 8, 12, 16, 20, 24,
and 28 times the rms noise.
\label{fig:P-V arm}
}
\end{figure}

Roughly 80\% of the BIMA CO emission from
NGC 5394 is concentrated in the central 800 pc (FWHM) starburst 
region,\footnote{The 800 pc size is from a Gaussian fit to the
$\lambda = 20$ cm radio continuum and \Halpha\ emission, deconvolved
from the point spread function.}
and the rest comes from the part of the inner disk  south and west of the
central starburst.
After subtracting the CO emission from the 800 pc starburst (assumed
axisymmetric), we get the residual $I_{\rm CO}$ image displayed in the top
panel of Figure~\ref{fig:residual}.
The innermost western arm, $6''$ west of the nucleus,
appears as an arm in H$\alpha$ with the brightest \Halpha\
emission along its inside edge (see bottom panel of Figure~\ref{fig:residual}).
Part of this arm is detected 
in CO emission, with a maximum $I_{\rm CO}$ of 31 K \kms, 
equivalent to a line--of--sight column density $N(\rm H_2)$ of
$7 \times 10^{21}$ H$_2$ cm$^{-2}$. The P--V diagrams in 
Figure~\ref{fig:P-V arm} are for a slice along the declination axis, $6''$
west of the nucleus, and thus display the emission from the innermost western
arm. These diagrams show that CO and \Halpha\ have a similar distribution 
of velocities as well as similar mean velocities at this arm. 
Since the \Halpha\ 
resolution is $2 \farcs 4$, the \Halpha\ P--V diagram in
Fig.~\ref{fig:P-V arm} contains negligible
emission from the starburst, whereas the CO P--V diagram contains
some starburst emission. 

Before correction for
the velocity gradient across the beam, the CO one--dimensional velocity
dispersion at the innermost western arm is 15 \kms. To correct for the 
velocity gradient,
we subtract the velocity difference across the Gaussian beam in quadrature.
We have confirmed with numerical experiments that subtracting in
quadrature is the appropriate correction (for plausible values of linear 
shear). This 
reduces the CO velocity dispersion to 14 \kms. 
Since estimates of the cloud--cloud CO velocity 
dispersion in the Milky Way range from 3 to 9 \kms, with a median value of
4 \kms\ \citep{COM91}, the CO velocity dispersion
at the innermost western arm seems to be somewhat higher than normal for
galaxy disks. If attributed to unresolved in--disk streaming motions, such
motions would need to be large because the disk is viewed nearly face--on
(inclination $i$ = 15\deg).
 Over much of the disk of NGC 5394, Paper I
found \HI\ velocity dispersions of 20 -- 40 \kms, which are high compared
to the \HI\ velocity dispersions of 6 -- 13 \kms\ in undisturbed spirals
(see references in \citet{KAU97}). 
Given the nearly face-on orientation of NGC 5394, these
velocity dispersions probably represent mainly the $z$--component of the
motions.

Paper I reported a weak  ($3 \times$ rms noise) detection in the Onsala 
observations of CO emission
from the southern tidal arm of NGC 5394, with $I_{\rm CO}$ = 3.3 K \kms.
The BIMA observations detect no emission here. 
As the $2 \times$ rms noise upper limit to  
$I_{\rm CO}$ in the BIMA data, we take $I_{\rm CO}$ = 
$2 \times$ (the rms noise per channel in the BIMA cube) $\times$
 (the width of two channels)/PB, where PB is the primary--beam sensitivity
of the mosaiced data.  
If the CO emission uniformly fills the Onsala $33''$ aperture on the southern
tidal arm, it would be too faint to detect in the BIMA observations since
the $2 \times$ rms noise upper limit to
$I_{\rm CO}$ at this position is 7.1 K \kms\ in the BIMA map.

\section{The Eastern Inner--Disk Arm of NGC 5394}

The eastern and the outer of the two western inner--disk arms are both 
devoid of on--going star formation. No CO emission is detected from the
eastern inner--disk arm. This is probably also the case for the outer
of the two western inner--disk arms, but at our CO resolution, the situation
there is contaminated by CO emission from the innermost western arm. 
Therefore we concentrate on the eastern inner--disk arm but 
assume that the same logic applies to the outer of the two western inner--disk
arms. 

Why is there no evidence of star formation in progress at the eastern 
inner--disk arm? We consider first the surface density of gas.
\HI\ emission from this arm is present 
with a line--of--sight column density $N$(\HI) = $7 \times 10^{20}$ \atc2\ 
(measured with resolution = $11''$ = 2.6 kpc HPBW in Paper I). The 
$2 \times$ rms noise 
upper limit to $I_{\rm CO}$ at the eastern inner--disk arm in the BIMA
data is 7.6 K \kms.  With the ``standard'' value of $X$, this corresponds
to a $2\times$ rms noise
upper limit to $N(\H2)$ of $1.7 \times 10^{21}$ \H2\ cm$^{-2}$. 

\begin{figure}
\epsscale{0.54}
\plotone{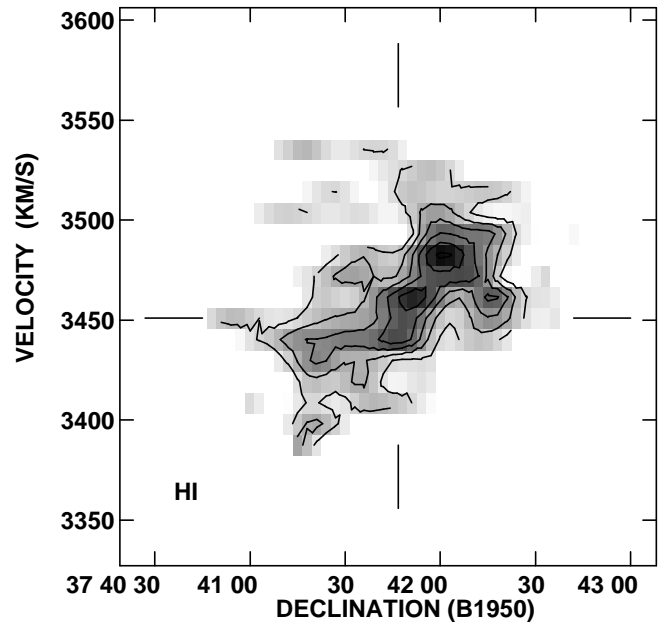}
\caption{Position--velocity diagram in \HI\ through the nucleus of 
NGC 5394  with the abscissa along the projection line--of--nodes of the
galaxy (position angle = 0\deg). The large ``cross--hairs'' mark the position
and velocity of the optical nucleus. The \HI\ contour levels are at 2, 3, 4, 5,
6, and 7 times the rms noise of 0.46 \mJybeam. The \HI\ diagram includes 
emission over a significantly greater declination range but about the same
velocity range as the CO emission in the P--V diagram in
Fig.~\ref{fig:P-V nucleus} (left panel). 
\label{fig:P-V HI}
}
\end{figure}

\begin{figure}
\epsscale{1.0}
\plottwo{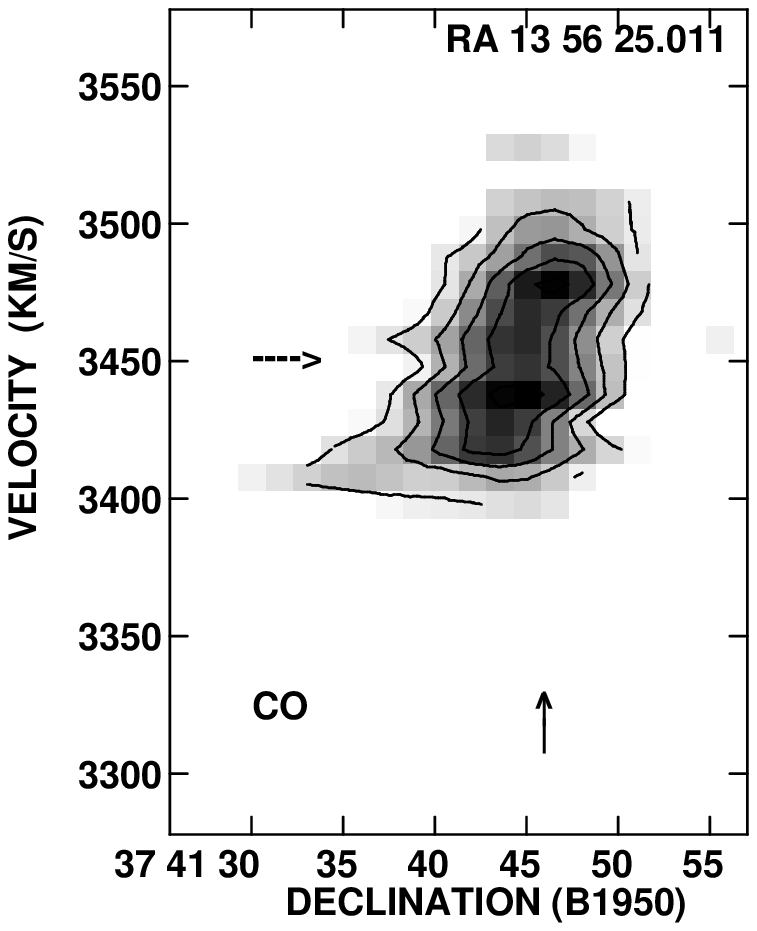}{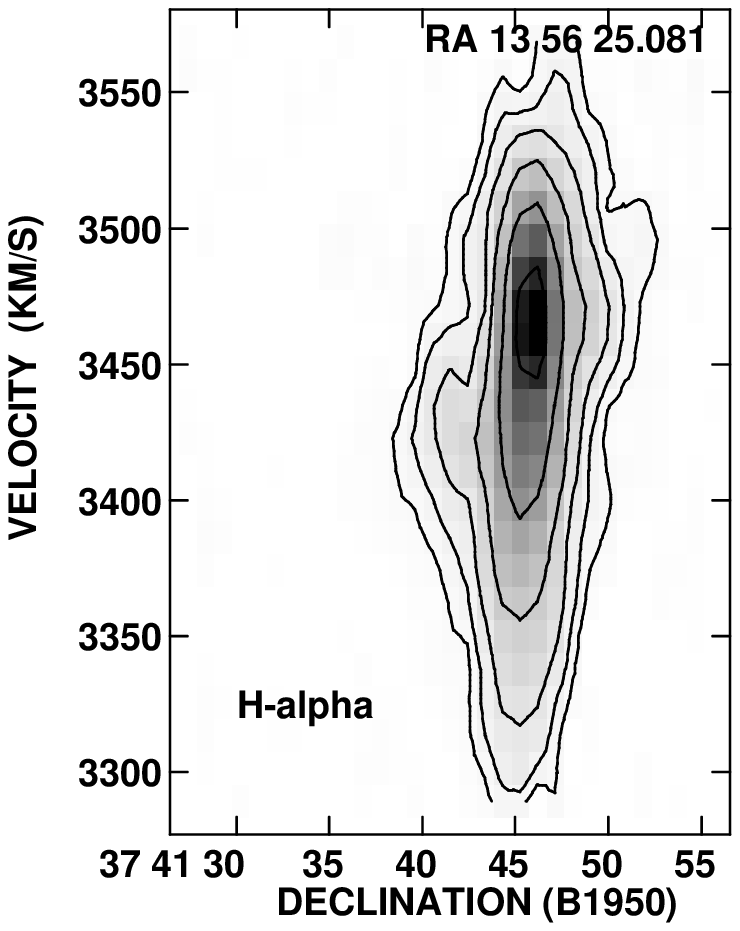}
\caption{Position--velocity diagram in CO (left panel) and \Halpha\ 
(right panel) through the nucleus of NGC 5394   
with the abscissa along the projection line--of--nodes of the
galaxy (position angle = 0\deg). The contour levels in CO are at 2, 4, 6, 8,
and 10 times the rms noise.  The contour levels in \Halpha\ are at 4, 8, 16,
32, 64, and 128 times the rms noise. In the CO P--V diagram, arrows mark the
declination of the nucleus and the long--slit \Halpha\ value for 
$v_{\rm sys}$. Unlike the \Halpha\ emission,
the CO emission from the center of NGC 5394 is double--peaked in velocity, 
with one maximum at 3478 \kms\ and the other at 3438 \kms.
An extension to low velocities in \Halpha\
at the nucleus is not seen either in CO or in \HI. 
\label{fig:P-V nucleus}
}
\end{figure}

From observations of spiral galaxies, \citet{KEN89} and \citet{MAR01} found that 
the instability parameter $Q_{\rm gas} \leq 1.4$
for significant star formation to occur.  
The instability parameter 
$Q_{\rm gas} = \kappa \sigma_{\rm v,gas}/\pi G \mu_{\rm gas}$, where $\kappa$
is the epicyclic frequency, $\sigma_{\rm v,gas}$ is the one--dimensional
velocity dispersion in the gas, and $\mu_{\rm gas}$ is the face-on surface 
density of gas. Paper I used \HI\ data to determine values for these 
parameters.
A position--velocity diagram in \HI\ from Paper I (with $18''$ resolution)
and a position--velocity diagram in CO 
(with $6 \farcs 6 \times 6 \farcs 3$ resolution), 
are displayed in Figure~\ref{fig:P-V HI} and in the left panel
of Figure~\ref{fig:P-V nucleus}, respectively,
for a cut through the optical nucleus with abscissa along the projection 
line--of--nodes of the galaxy (position angle = 0\deg). The \HI\ 
P--V diagram provides a rather messy ``rotation curve''
because NGC 5394 is viewed nearly face--on, 
the \HI\ velocity dispersion is high,
there are streaming motions associated with the tidal arms and the bar,
and the spatial resolution is low.
From the \HI\ data, Paper I concluded that the inner--disk arms are located
on the solid body part of the rotation curve and took the 
epicyclic frequency at galactocentric distance 
$R = 10''$ (on the eastern inner--disk arm) as
$\kappa = 2 \Omega$ = 37 \kms\ kpc$^{-1}$, where 
$\Omega$ is the orbital angular speed.\footnote{The value was misprinted in
Paper I as 3.7 \kms\ kpc$^{-1}$.} Also, the inclination $i$ was estimated
 to be 9\deg\
by comparing the observed value of $v_{\rm max} \sin i$ = 27 \kms\ with the 
expected value of $v_{\rm max}$ = $163 \pm 26$ \kms\ from 
the $I$--band luminosity (see Paper I).

 With the higher angular resolution of
the CO P--V diagram, we see that the rotation curve has a steeper rise in
the inner disk than apparent from the \HI\ data. It may have turned 
over by $R = 10''$ in the south (see Figure~\ref{fig:P-V nucleus}), but
it is difficult to draw a firm conclusion because the
emission at this radius is faint, the velocities are affected by the streaming 
along the bar, and the CO resolution is $\sim 6.5''$. At $R= 10''$
we find $\Omega$ = 41 \kms\ /(2.3 kpc $\sin i$). The 
added information from the CO P--V
diagram changes $v_{\rm max} \sin i$ to 41 \kms, and thus changes
our estimate of the inclination $i$ to 15\deg\ and
our estimate of the epicyclic frequency
$\kappa$ to
[2 (41 \kms)]/(2.3 kpc $\sin i$) = $1.4 \times 10^2$ \kms\ kpc$^{-1}$
at $10''$ if in solid body rotation, or a factor of 
$\sqrt 2$ less if on the flat part of the rotation curve. 

Since the distribution of gas is lopsided, we use the surface density of
gas on the eastern inner--disk arm at $R = 10''$ to
compute the instability parameter. For $\sigma_{\rm v,gas}$, we adopt 
from Paper I the
\HI\ value of  30 \kms\ at the eastern inner--disk arm. The 
\HI\ column density at this arm is $N$(\HI) = $7 \times 10^{20}$ \atc2. 
Including a 36\% by mass contribution from helium, 
$$
\mu_{\rm gas} = (1 + 0.36) [N(\rm {HI}) + 2 N(\H2)] m_{\rm p} \cos i,
\eqno(1)
$$
where $m_{\rm p}$ is the proton mass. (The calculation of $\mu_{\rm gas}$
in Paper I ignored the contributions from helium and molecular gas.)
Using the above $2\times$ rms noise upper limit to $N(\H2)$ gives an upper
limit to $\mu_{\rm gas}$ of 43 \Msolpc2\ at $R = 10''$ on the eastern 
inner--disk arm and a lower
limit to $Q_{\rm gas}$ of 5.0 if on the 
flat part of rotation curve.  An upper limit to $Q_{\rm gas}$ is
obtained by setting $N(\H2)$ = 0: $Q_{\rm gas}$ = 29 
if on the flat part of the rotation curve. The values of $Q_{\rm gas}$
should be increased by $\sqrt 2$ if in solid--body rotation.  
Thus $Q_{\rm gas}$ lies in the range 5 to 42. 
Since this exceeds critical value of 1.4 from \citet{KEN89} (see also
\citet{MAR01}),
it appears that the instability parameter
$Q_{\rm gas}$  is presently too high for
significant star formation at the eastern inner--disk arm. 

For comparison, star formation is presently occurring on the  innermost 
western arm and CO emission is detected from a portion of that arm. To
compute $Q_{\rm gas}$ there, we take our peak value of $I_{\rm CO}$ at
this arm, which corresponds to an $N(\H2)$  of 
$6.9 \times 10^{21}$ \H2\ cm$^{-2}$ and use our measured CO velocity
dispersion of 14 \kms\ for $\sigma_{\rm v,gas}$ since $N(\H2)$ dominates
$N(\HI)$ here. Then with solid body rotation, we get $Q_{\rm gas} \sim 1$.
So the Toomre gravitational instability criterion seems to explain the
presence of star formation on the innermost western arm, provided the
value of the conversion factor $X$ is not appreciably less than the
standard value.  

The simple answer to the question of why there is no evidence of ongoing star 
formation at the eastern inner--disk arm or at the outer of the two
western inner--disk arms is that there is presently too
little molecular gas there. 
The question becomes why is there so little molecular
gas associated with optically bright spiral arms.

\section{Central Starburst in NGC 5394}

\subsection{CO Flux, Star Formation Rate, and Extinction}

The total $S_{\rm CO}$ in NGC 5394 detected by BIMA is 117 Jy \kms. Of this,
approximately 97 Jy \kms\ is from the starburst region (on the assumption of 
east--west symmetry for the starburst). 
Using the adopted standard value of $X$, we make the following comparison
between the NGC 5394 starburst and the M82 starburst. The NGC 5394 starburst
has a radio continuum luminosity 
at $\lambda = 20$ cm of $7.7 \times 10^{21}$ W Hz$^{-1}$ (Paper I), a linear 
size of 800 pc, and an H$_2$ mass $M(\rm H_2)$ of 
$1.9\times 10^9$ \Msol\ (excluding helium). The M82 starburst has a radio 
continuum luminosity of $1.0 \times 10^{22}$ W Hz$^{-1}$, a size of 600 pc
\citep{CON90}, and an $M(\rm H_2)$ measured
by \citet{WIL92} 
for a region with diameter 800 pc and scaled to the same value of $X$
of $4.9\times 10^8$ \Msol. 
Thus, compared to the M82 starburst, the starburst in NGC 5394
has about the same linear size, 77\% of the $\lambda 20$ cm radio continuum 
luminosity, but $\sim 4$ times the CO luminosity. It may be that the M82
starburst is in a more advanced state and has already used up more of the
gas. \citet{RIE93} suggested that the M82 starburst 
consists of two 
starbursts, one of which started $3 \times 10^7$ years ago and the other, 
$5 \times 10^6$ years ago. \citet{VAN98} found that 
the starburst in NGC 5394 is several Myr past the peak star--formation rate. 

For the position of the nucleus, we take the location of the bright
radio continuum peak in the FIRST image. (The stated uncertainty of
the FIRST survey position is $0 \farcs 4$). 
There is negligible displacement between the surface brightness maxima
in the radio continuum, $J$--band, and \Halpha.
The $I_{\rm CO}$ maximum is $0 \farcs 8$ west and $0 \farcs 4$ south
of the radio continuum position. Since
the positional uncertainty of the CO maximum is less than $0 \farcs 6$,
the offset between the $I_{\rm CO}$ and radio continuum maxima is
marginal. It could
result from some depletion of the CO flux at the radio continuum maximum, 
e. g., by photodissociation of CO or by star formation consuming the gas.
Gaussian fits to the
emission (Table 3) show that the \Halpha\ source and the radio
continuum source in the center of NGC 5394 have the same size, 800 pc (FWHM).
The CO emission with a major axis of 1.1 kpc (FWHM) is slightly more
extended. The radio continuum emission from NGC 5394 is nonthermal
with a spectral index of $-0.7$,  and 83\% of
the  radio continuum flux density $S_\nu(20)$ of NGC 5394 at 
$\lambda = 20$ cm is from the starburst region (see Table 1).

 There is no evidence that the emission
from the center of NGC 5394 is dominated by an AGN. The optical and 
near--infrared spectra \citep{LEE89, VAN98} 
 are \HII\ region--type. The radio
continuum emission does not appear to be sufficiently compact to be mainly 
from an AGN. It is likely that 
most of the radio emission originates from SNRs or 
cosmic--ray electrons previously accelerated in supernovae, as is
the case for the M82 starburst.   

The measured \Halpha\ flux from the NGC 5394 starburst, not corrected for
extinction, implies a radio
continuum flux density from optically thin free--free emission 0.9\%
of the measured value of $S_\nu(20)$ (see Paper I).
We can get a firm upper limit to the average visual extinction 
$A_{\rm v}$ of the \HII\ regions by assuming that all 
the measured $S_\nu(20)$ is free--free
emission (clearly it is not, since the spectral index is nonthermal).
The resulting  upper limit,
 $A_{\rm v}$ = $1.28 \times A(\Halpha)$ = 6.5 mag, is unrealistically high
since it assumes a thermal fraction at $\lambda$ 20 cm of 100\%. For a better
estimate we choose the following two alternatives.
In the M82 starburst, the thermal fraction at $\lambda$ 20 cm is
7\%, as determined by \citet{GOL96} 
from a comparison of [Ne II] 12.8 \micron\ and radio continuum emission.
If the same thermal fraction applies to the NGC 5394
starburst, then the intrinsic \Halpha\ flux is a factor of 7.5 greater
than the measured \Halpha\ flux, and the average extinction for star--forming
regions in the NGC 5394 central starburst is $A_{\rm v}$ = 
$1.28 (2.5) \log 7.5$ = 2.8 mag. If, instead,
we adopt the $\lambda 20$ cm thermal fraction of $\sim 13\%$ typical of
normal disk galaxies \citep{CON92}, then the 
intrinsic \Halpha\ flux is a
factor of 14 times the measured \Halpha\ flux, and the average extinction
for the \HII\ regions is $A_{\rm v}$ = 3.7 mag. 

Attributing all of the $IRAS$ 60 \micron\ flux of the Arp 84 system to 
NGC 5394, \citet{LEE89} derived  a luminosity ratio $L(60 \micron)/L(\Halpha)$
of 2000 for the NGC 5394 starburst. For the disks of bright 
spiral galaxies, \citet{LEE89} found $L(60 \micron)/L(\Halpha)$ equals 
20 -- 200 (based on data from \citet{PER87} in which the \Halpha\ flux of the
disk was corrected for a mean extinction of 1.1 mag at \Halpha).
With a resolution of $4.8' \times 1.5'$ at 60 \micron, the $IRAS$ 60 \micron\
flux of Arp 84 includes both galaxies. To determine what fraction of the 
$IRAS$ flux is from the NGC 5394 starburst, we use the FIR--nonthermal radio
correlation for spiral galaxies. Since Paper I finds that 29\% of the radio
continuum flux from the galaxy pair is from the central starburst in
NGC 5394, we attribute 29\% of the $IRAS$ 60 \micron\ flux to the NGC 5394
starburst. This reduces the value of $L(60 \micron)/L(\Halpha)$ for the  
starburst to $\sim 600$. If we adopt the range of values (2.8 -- 3.7 mag) 
for the average A$_{\rm v}$ from the thermal fraction
argument, then the value of
$L(60 \micron)/L(\Halpha)$ is further reduced to the range 40 -- 80, similar
to values found in the disks of bright spirals. Adoption of an average 
$A_{\rm v}$ of 3 -- 4 mag gives a plausible consistency between 
the $IRAS$ 60 \micron\ flux, the radio continuum, and the \Halpha\ flux.
It thus appears that the intrinsic \Halpha\ flux is an order of magnitude
greater than the observed \Halpha\ flux. In that case, the
dereddened \Halpha\ luminosity of the
starburst region is  $\sim 8 \times 10^{41}$ erg s$^{-1}$.
Then, taking the star formation rate 
SFR = $7.9 \times 10^{-42} L(\Halpha)$ \Msol\ yr$^{-1}$ from
\citet{KEN98}, we get an SFR of $6 \pm 2$ \Msol\ yr$^{-1}$ 
for the starburst region. 

The \Hbeta\ to \Halpha\ emission--line ratio (Balmer decrement)
measured for the NGC 5394 nucleus by \citet{LEE89} with 
a $1 \farcs 5$ wide slit gives a visual extinction $A_{\rm v}$ of 1.3 mag
(assuming Case B recombination). They find broad--band
optical colors, uncorrected for reddening, similar to those of G stars.
They note
that the \Halpha\ flux is roughly consistent with the number of ionizing
stars implied by the broad--band optical fluxes. 
In the BIMA data, the maximum $I_{\rm CO}$ in the nucleus is
$1.7 \times 10^2$ K \kms, 
and the mean value of $I_{\rm CO}$ for the starburst
is 66 K \kms. With the standard value of the conversion factor $X$, the maximum 
$I_{\rm CO}$ implies a column density 
$N(\H2) = 3.8 \times 10^{22}$ \H2\ cm$^{-2}$, and the mean $I_{\rm CO}$,
a column density $N(\H2) = 1.5 \times 10^{22}$ \H2\ cm$^{-2}$.
For a standard dust--to--gas ratio with $A_{\rm v}/N_{\rm gas}$ =
$0.53 \times 10^{-21}$ mag (\atc2)$^{-1}$ \citep{BOH78}, the
mean CO column density implies a highly opaque region with
$A_{\rm v}$ = 11 mag to the midplane. Although it is known that the 
dust--to--gas ratio is proportional to metallicity $Z$ 
\citep{SOD95}, we lack a reliable determination of the metallicity 
of the NGC 5394 starburst. \citet{LEE89} note that the measured value of
the [\ion{N}{2}]/\Halpha\ emission--line ratio for the starburst is compatible
with solar metallicity but warn this is not conclusive
because the ionization parameter is unknown.

How can we reconcile these three very different values for $A_{\rm v}$?
The small value of $A_{\rm v}$ (1.3 mag) from the Balmer decrement as compared
to the average extinction (3 -- 4 mag) from the thermal fraction argument 
means that the observed optical flux originates from regions of small 
optical depth (near the edges of the clouds),
and we are not seeing the optical photons emitted by stars and
\HII\ regions deeply embedded in opaque clouds. Thus, optically, most of the 
starburst is hidden from view,
probably in a patchy distribution of highly opaque clouds, as 
\citet{LEE89} suggest. 

The value of $A_{\rm v}$ (11 mag) deduced from the mean $I_{\rm CO}$ with 
standard values for $X$ and the dust--to--gas ratio exceeds 
the average $A_{\rm v}$ of 3 -- 4 mag from the radio thermal fraction versus
\Halpha\ and $L(60 \micron)/L(\Halpha)$ arguments. The most likely 
explanation is that the conversion factor $X$ = $N(\H2)/I_{\rm CO}$ in 
the starburst region is 
a factor of 3 -- 4 below the standard value. Similar 
reductions in the value of $X$ 
have been obtained for high excitation regions in the centers of other galaxies
\citep{WAL93,REG00}. For the Milky Way, 
\citet{SOD95} concluded that $X$ is a factor of 
3 -- 10 lower near the Galactic center than in the disk at 
R = 2 -- 7 kpc. 
In the M82 starburst, \citet{SMI91}  and 
\citet{WIL92} found that 
the value of $X$ is a factor of 3 below the standard value.
 It appears that the different physical conditions in a starburst nucleus
\citep{BRO93,AAL95,AAL97} result in a smaller value of $X$ than
in the main disk of a spiral galaxy. \citet{KWH99} calculate the dependence of
$X$ for a photodissociation region on far--ultraviolet flux, density, 
column density, and metallicity and apply their model to the M82 
starburst as an example. They can fit the M82 observations with clouds that
have higher gas density and higher thermal pressures than GMCs in the
Milky Way.  

Use of the standard
value of $X$ overestimates the extinction in the starburst region of NGC 5394
 and, consequently, overestimates the total molecular mass of NGC 5394.
If the value of $X$  is down by a factor of 3 -- 4 in the starburst region
but not reduced outside of the starburst, then the total mass in gas in the
galaxy is $(3.6 \pm 0.4) \times 10^9$ \Msol.

\subsection{Velocity Structure}

Figure~\ref{fig:P-V nucleus} displays position--velocity diagrams 
through the nucleus of NGC 5394 in
CO and in  H$\alpha$ (from the Fabry--Perot data of Paper I)
with the abscissa along the projection line--of--nodes of the
galaxy (position angle = 0\deg).
In the Fabry--Perot \Halpha\ data, the mean velocity at the
nucleus is 7 \kms\ greater than the long--slit value of $3451 \pm 12$ \kms\
obtained by \citet{KAR80}; this indicates the uncertainty
in the calibration of the Fabry--Perot \Halpha\ velocities at the 
NGC 5394 nucleus. The channel width is 12 \kms\ in the \Halpha\ cube 
and 10 \kms\ in the CO cube.  These P--V diagrams reveal two significant 
differences between the CO and H$\alpha$ velocity distributions in the
starburst region: (a) the CO emission is double--peaked in velocity whereas
the \Halpha\ emission has a single velocity peak and (b) in \Halpha\
a clear extension to lower velocities is seen at the nucleus but is
absent in CO. 
The CO emission has one maximum at 3478 \kms\ and the other 
at 3438 \kms; the average of these two velocity peaks, 3458 \kms, is
close to the \Halpha\ systemic velocity.
The separation in velocity implies a spatial separation, e.~g., one velocity
peak from the approaching side and the other from the receding side.
 This could result
from some depletion of CO at the nucleus by photodissociation or 
consumption by star formation or
the molecular gas could be in a nuclear ring or in
``twin peaks'' near an inner Lindblad resonance (ILR).
It also means that the two CO peaks are not spatially coincident with the
peak in the \Halpha\ emission.  If the double peaks in CO are in the 
disk plane of NGC 5394, then each has a velocity difference with respect to 
the nucleus of $(v_{\rm obs} - v_{\rm sys})/\sin i$ = 20 \kms/$\sin 15\deg$ = 
77 \kms. 

We made CO position--velocity diagrams along various position angles.
The maximum separation between the two CO peaks is $\sim 2''$ 
(less than 1/3 the synthesized beamwidth), with a smaller
spatial separation along the bar major axis than along the bar minor axis.
It is only the
velocity structure that enables us to identify these as separate features.

\begin{figure}
\epsscale{1.0}
\plottwo{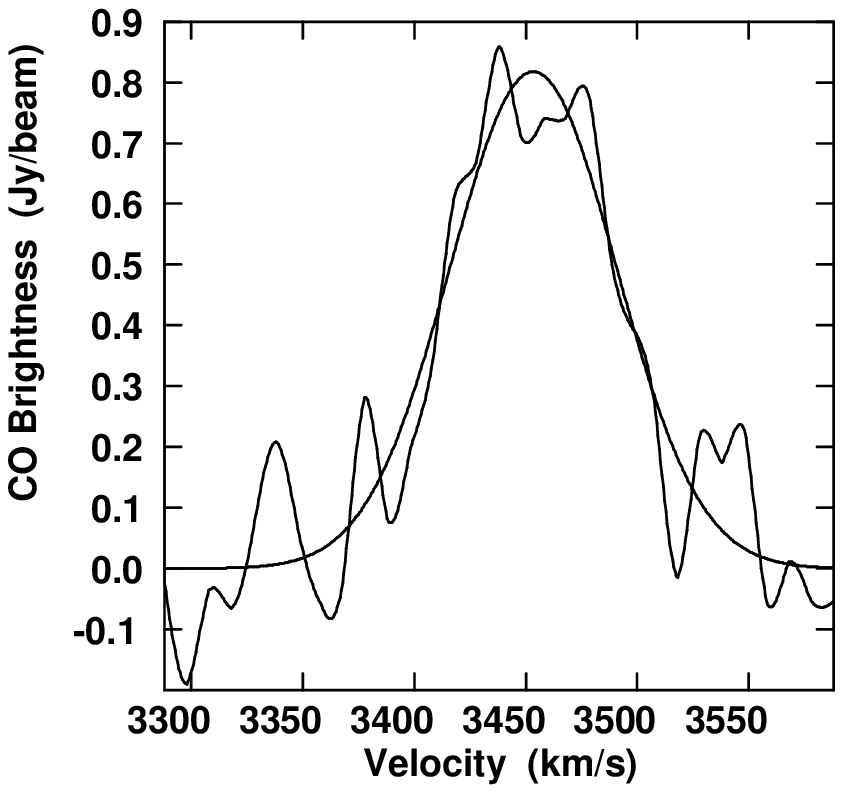}{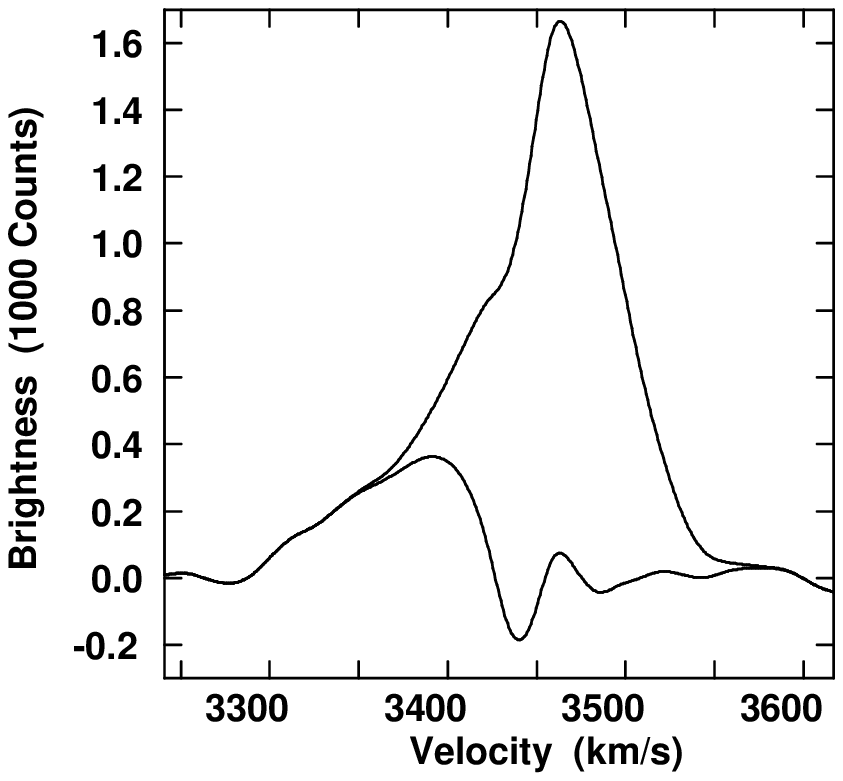}
\caption{Line profiles in CO (left panel) and
\Halpha\ (right panel) at the nucleus of NGC 5394.
The Gaussian fit to the CO line profile and
the residual spectrum after fitting the \Halpha\ line profile 
(as described in the text) are also displayed. The dip at 3450 \kms\
in the CO profile occurs close to the long--slit \Halpha\ value of
3451 \kms\ for $v_{\rm sys}$. 
\label{fig:nucleus profile}
}
\end{figure}

Figure~\ref{fig:nucleus profile} compares the CO line profile at the nucleus 
with the \Halpha\ line profile from Paper I.
From a Gaussian fit to the 
CO line profile at the nucleus, we find that the CO central velocity is
3453 \kms\ at the $I_{\rm CO}$ maximum and that the FWHM is 
$88 \pm 2$ \kms. Paper I 
suggested that the extended blue wing of the nuclear \Halpha\ profile
(a velocity asymmetry of $\sim 100$ \kms) 
could represent an outflow of ionized gas from the starburst, perpendicular
to the disk. From the above extinction estimate, a
counterflow on the far side would not be visible in \Halpha. 
On the assumption that the \Halpha\ nuclear profile consists of
a symmetric Gaussian component from the starburst in the nuclear disk plus 
an outflow/wind (visible on the near side only),
we fit the high velocity side of the profile with one side of a 
Gaussian with FWHM = 77 \kms\ and then subtracted the (full)
Gaussian profile to 
obtain the residual \Halpha\ spectrum shown in Fig.~\ref{fig:nucleus profile}
for the outflow.
In the \Halpha\ profile, the maximum residual intensity in the blue wing
(which we associate to an outflow) is 22\% of the maximum in
the original line profile.  
There is a hint of a feature ($ 2 - 2.5 \times$ the rms noise)
in the CO profile at 80 -- 100 \kms\ to the red of $v_{\rm sys}$,
but no clear evidence for a molecular counterpart to the \Halpha\ 
outflow component. 

For comparison, in
M82 the optical emission lines give an outflow speed of $\sim 600$ \kms\
\citep{SHO98}, whereas the \12CO\ and higher level CO molecular lines
give an outflow speed of only $\sim 200$ \kms\ 
\citep{SHE95,NAK87,SEA01}.
Also, in M82, the line splitting that occurs in \Halpha\
is not seen in CO. If in the NGC 5394 starburst, the ratio of \Halpha\
to molecular outflow speeds is similar to that in M82, then 
in NGC 5394 it would be hard to identify
an outflow in CO (that is, to distinguish between an outflow and a thick disk 
or spheroidal distribution)
 since the \Halpha\ outflow speed is only $\sim 100/\cos 15\deg$ \kms. 

The CO data have lower spatial resolution than the \Halpha\ data. If we
correct for the gradient of the mean velocity field across the respective
beams (by subtracting the velocity difference across the beam in quadrature), 
then the CO profile at the nucleus has a one--dimensional velocity 
dispersion of 31 \kms\  and the 
part of the \Halpha\ line profile that we attribute to 
the nuclear disk has a one--dimensional velocity dispersion of 30 \kms. 
This probably represents mainly the $z$--component of the velocity dispersion. 
The molecular and \Halpha\ velocity dispersions at the 
NGC 5394 nucleus are about the same size as the $z$--component
of the molecular velocity dispersion (30 -- 54 \kms) in the Milky Way
Galactic Center estimated by \citet{BAL88}. The smaller value for the 
molecular velocity
dispersion found in Section 4 pertains to the innermost western arm, not
the nucleus.

\section{Comparison of CO, \Halpha, and \HI\ Velocity Fields of NGC 5394}

\begin{figure}
\epsscale{0.7}
\plotone{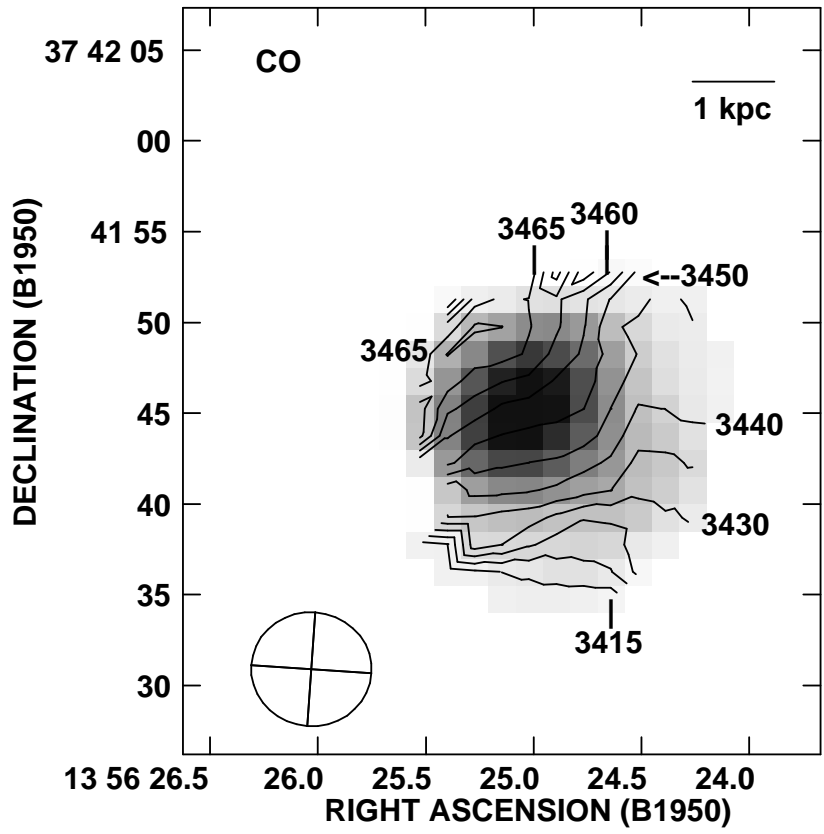}
\caption{CO isovelocity contours overlaid on the CO integrated intensity image
of NGC 5394 in grayscale.
\label{fig:CO velocity}
}
\end{figure}

\begin{figure}
\epsscale{0.7}
\plotone{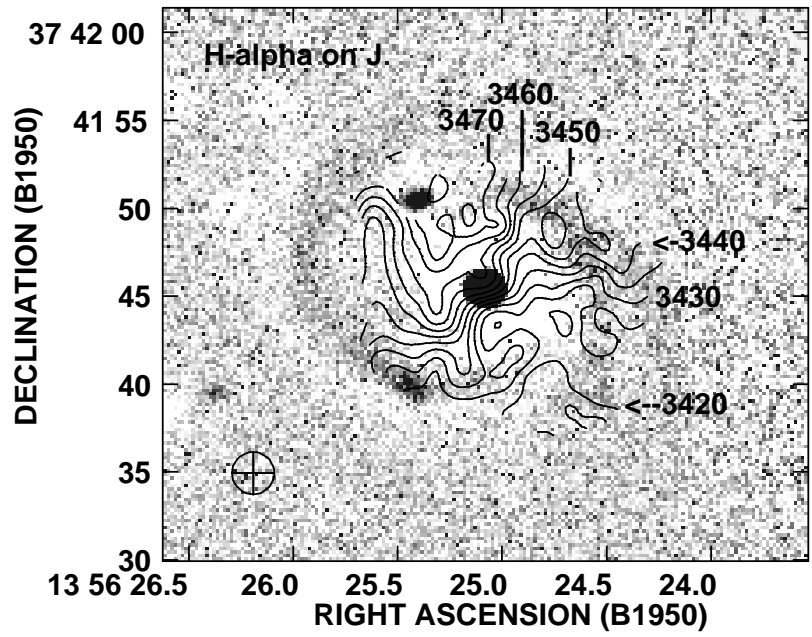}
\caption{\Halpha\ isovelocity contours overlaid on  unsharp--masked 
$J$--band image of NGC 5394
in grayscale. The beam symbol represents the \Halpha\ resolution.
\label{fig:Halpha velocity}
}
\end{figure}

Figure~\ref{fig:CO velocity} displays the \12CO\ velocity field from the 
BIMA data, and Figure~\ref{fig:Halpha velocity},
the \Halpha\ velocity field (from Paper I). The CO velocity field covers only 
$\sim 2.5 \times$ the BIMA synthesized half--power beamwidth and does not
reveal as much detail as the \Halpha\ velocity field. 
Given the difference in spatial resolution, the CO velocity field appears 
quite similar to the \Halpha\ velocity field. The similarity, despite 
\Halpha\ extinction, can be understood if the \Halpha\ and CO emitting
layers in the disk have negligible shear perpendicular to the disk. 
The more extended VLA \HI\ data in Paper I
($18''$ resolution) provide the velocity field of the main disk and the 
tidal arms.  In the main disk, the \HI\ velocity field has kinematic minor 
axis at $90\deg \pm 2\deg$, so Paper I took 
the position angle of the galaxy's projection line of nodes as 
$0\deg \pm 2\deg$. The near side of the galaxy is the western side.
In the CO and \Halpha\ velocity fields,
the kinematic minor axis has rotated such that in the central
$10''$ it has a very different position 
angle of $125\deg \pm 5\deg$ and is along the major axis of the 
$16''$--long primary bar. Some possible explanations for the rotation of a 
kinematic minor axis in the
inner disk of a galaxy are streaming along a bar or  
axisymmetric radial motions. The sense of rotation of the kinematic 
minor axis of NGC 5394 is opposite to that
expected for axisymmetric radial inflow in the disk. Axisymmetric
radial expansion in the disk  would rotate 
the kinematic minor axis in the sky--plane by 
$\tan \Delta \phi_{\rm minor} = - (v_{\rm exp}/v_{\rm t}) \sec i$, where 
$\Delta \phi_{\rm minor}$ equals the  position angle of the minor axis
of the projection minus the position angle of the kinematic minor axis of
the central region, $v_{\rm exp}$ is the expansion speed 
(positive for expansion),  and
$v_{\rm t}$ is the tangential speed \citep{WON00}.
Since the observed $\Delta \phi_{\rm minor}$ =  $90\deg - 125\deg$
= $-35\deg$ is negative, there is no evidence that axisymmetric inflow of 
gas into the 
inner disk is presently occurring. If axisymmetric expansion, the implied
value of $v_{\rm exp}$ is large, 0.68 $v_{\rm t}$.
The more likely interpretation
of the rotation of the kinematic minor axis is that it represents
gas moving in elliptical orbits around the bar. 
The inner--disk arms may have a different pattern speed than the primary
bar, as they do not
depart from the ends of the bar. 

Since NGC 5394
is viewed nearly face--on, large in--plane velocity excursions would not
appear very prominent in any line--of--sight velocity field. With its
higher spatial resolution, the \Halpha\ velocity contours have more wiggles,
indicative of streaming motions, than the CO velocity contours.
Some of these appear associated with the innermost western arm. 
Additionally, it seems likely 
that the $u$--shaped kinks in the \Halpha\ velocity contours $3''$
southeast of the nucleus and the $n$--shaped velocity kinks $3''$ 
northwest of the nucleus are related to the possible secondary bar 
visible in the $J$ and $K$--band images in 
\citet{BUS92}.
The bilateral symmetry of the velocity kinks at $3''$ provides kinematic 
evidence that this stellar feature is a coherent structure produced by
a non--axisymmetric gravitational potential.
The position angle of the kinematic minor axis changes from 
$125\deg \pm 5\deg$ at $R \sim 5''$ to $102\deg \pm 5\deg$ at $R \sim 2''$;
the kinks represent the transition between these two orientations. 
 Thus the velocity field
supports the interpretation that there are two nested bars, with the
secondary bar trailing the primary bar by 30\deg. It is
puzzling that the secondary bar is not seen in the higher resolution $J$--band
image of Paper I.

\section{Discussion of Results on NGC 5394}

The BIMA \12CO\ observations of the tidally--distorted galaxy NGC 5394
reveal that:
(1) about 80\% of the CO emission detected by BIMA is concentrated in the 
800 pc (FWHM) starburst region; (2) the extended CO and  \Halpha\ 
emission in the inner disk have a similar lopsided spatial distribution, 
with an excess on the southwestern side of the disk, so star 
formation in the disk appears to be following the gas distribution; 
(3) no CO or \Halpha\ emission are detected
from the bright, eastern inner--disk arm; (4) the estimated value of
$X$ in the starburst is a factor of 3 -- 4 below the standard value. 
The \HI\ emission is also most
prominent on the same (western) side of the disk as the CO and 
\Halpha\ emission, so in total gas NGC 5394 has a lopsided spatial
distribution in the disk. 

Reducing the value of the conversion factor $X$ by a factor of 3 -- 4
in the starburst region but using the standard value of $X$ outside of
the starburst, we find from the Onsala $S_{\rm CO}$ that
$M({\rm H_2})/M(\rm HI)$ = 2.5 -- 2.7, 
 and that 20\% -- 25\% of the atomic plus molecular gas in the 
galaxy is located in the central starburst. 
 Galaxies with this high a value for $M(\rm H_2)$/$M$(HI)
are uncommon in the large sample of spiral 
galaxies surveyed by \citet{CAS98}. 
Using the standard value for $X$, they find the mean value 
of $M({\rm H_2)}/M(\rm HI)$ is 0.28 for Sa to Sc galaxies.
Like other starburst galaxies, NGC 5394 is CO--rich.
Also, using the standard value of $X$, \citet{CAS98} 
find that the average 
values of $[M({\rm H_2)} + M({\rm HI)}]/D_{25}^2$
range from $5.6 \times 10^6$ \Msol\ kpc$^{-2}$ for Sa galaxies
to $1.0 \times 10^7$ \Msol\ kpc$^{-2}$ for Sd galaxies, where $D_{25}$ 
is the isophotal diameter. They emphasize that   
$[M({\rm H_2)} + M({\rm HI)}]/D_{25}^2$ 
is simply a means of normalizing the gas mass to
a measure of the galaxy size and should not be interpreted as a gas surface
density. Since NGC 5394 is intrinsically oval, we set the value of $D_{25}$
for NGC 5394 equal to 
$2 (R_{25})_{\rm eff}$, where $(R_{25})_{\rm eff}$ is defined in Section 3.
Reducing the value of $X$ by a factor of 3 -- 4
in the starburst region but using the standard value of $X$ outside of
the starburst gives 
$[M({\rm H_2)} + M({\rm HI})]/D_{25}^2$ =
$(1.2 \pm 0.1)  \times 10^7$ \Msol\ kpc$^{-2}$, which is typical of late--type 
spirals. The present lack of star formation in two of the three inner--disk
arms results not from an overall paucity of gas in NGC 5394 but from its
spatial distribution.  

\begin{figure}
\epsscale{0.54}
\plotone{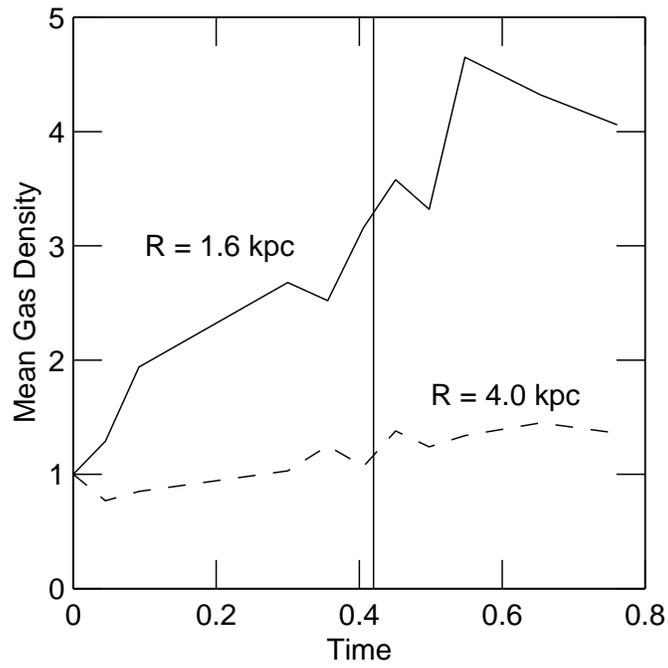}
\caption{Model results on central densities of NGC 5394 as a function of
time from the galaxy encounter simulation of Paper I. The curves represent
the ratio of the mean density to the initial density for
a spherical volume of radius 1.6 kpc (solid curve) and for a spherical
volume of radius 4.0 kpc (dashed curve).
The vertical line marks the closest approach time, while the present time
is near the latest time shown. The time unit is 
$4.2 \times 10^7$ years.
\label{fig:model}
}
\end{figure}

Paper I presented a galaxy encounter simulation that reproduces some of the
main features of this galaxy pair with a collision that is prograde relative
to NGC 5394 and retrograde at a high tilt angle relative to NGC 5395.
In the model, the response of NGC 5394 to the $\cos 2 \theta$ tidal 
potential of the encounter produces an eye--shaped (``ocular'') oval,
 which subsequently evolves into inner--disk arms, like those
visible in NGC 5394. There is a west versus east time lag in the
development of the ocular oval, with the eye--shaped rim forming first
on what is now the eastern side of NGC 5394. The model keeps count
of the number of gas particles in spherical volumes of radii 1.6 kpc and
4 kpc in the center of NGC 5394 as the encounter progresses. 
This is displayed in Figure~\ref{fig:model}. The limited particle resolution 
of the model prevents us from treating smaller volumes. 
 The number of gas particles within 1.6 kpc of the center 
builds up fairly steadily so that by the time of closest approach the
mean gas density of this region is three times its initial value and by the
present time it has leveled off at four times its initial value. The model
finds that the amount of material in the central regions is no longer
increasing at the present time; recall that the observed velocity field shows 
no evidence of present axisymmetric inflow (see Section 7). 

The dominant tidal perturbation to NGC 5394 is an $m = 2$ mode. This drives 
angular momentum transport, which produces the long tidal tails and inflow
into the central region. The
encounter also involves an $m = 1$ term, which
accounts for the time lag in the development of the eastern and western 
sides of the ocular oval. 
The model for this system (described in Paper I) uses rigid halos (softened
point--mass potentials). As the galaxies pass each other in the model, 
their disks do not quite interpenetrate but their halos do. This results
in a transient $m = 0$ compression in the perturbation: 
before closest approach, NGC 5394 compresses as a result of the increasing 
tidal force of the NGC 5395 halo. This contributes to the infall into the
central region before closest approach, while after closest approach the inflow
into the central region results from
ocular--induced angular momentum transport and inertia.

The observed lopsided distribution of gas could result from 
an asymmetry in the collisional perturbation plus  the loss of 
angular momentum by the gas in shocks. The latter is necessary
to differentiate between the gas and the stars because the distribution of
old stars does not show the same lopsidedness as the gas.
In the model, the
asymmetry of the perturbation is evident as a time lag in the 
development of the two sides of the ocular oval. Since the stars form a
collisionless system, they may respond differently than the gas, which
can lose angular momentum in shocks. We offer the following speculative
scenario for the inner--disk arms.
The west versus east time lag in the development of the
ocular oval may have led to a west versus east time lag in the formation of the
inner--disk arms and to a west versus east time lag in the inflow of gas to
the center. Gas at the eastern and the outer of the two western inner--disk 
arms underwent shocks (e.g., as a result of the bar, which may differ in 
pattern speed from the inner--disk arms) and flowed towards the center,
leaving the stellar arms behind.
 This needs to be confirmed with more detailed simulations.
Another possibility is that  there may have been more gas
inflowing towards the center on the western side.

Not all interacting galaxies are CO--rich \citep{HOR97}.
From a comparison of the observed morphologies with models for prograde,
in--plane, grazing encounters, Paper I concluded that the galaxies IC 2163,
NGC 2535, and NGC 5394 form three distinct stages in the development of
structures in this type of collision, with IC 2163 the least evolved and
NGC 5394 the most evolved interaction. The ``age'' sequence is also
seen in the star formation rates. IC 2163 has star formation typical 
of normal spiral disks \citep{ELM01}, and 
widespread CO emission
is found in SEST observations with $44''$ resolution (Thomasson et al.,
in progress). In NGC 2535, star formation is somewhat enhanced
\citep{BER93}, the brightest \Halpha\ emission \citep{AMR89} 
occurs at the eastern and western ends of the eye--shaped oval 
at $R \approx 5$ kpc from the nucleus, 
and the Onsala observations with $33''$ resolution 
had clear detections of CO emission only from the central pointing
(which includes the western apex) and from the portion of the tidal tail
just beyond the eastern apex \citep{KAU97}. Both positions have
about the same integrated intensity $I_{\rm CO}$. With the standard
value of $X$, NGC 2535 has $M(\rm H_2)$/$M$(HI) = 0.1. In NGC 5394,
most of the CO emission is from the central 800 pc starburst region, and
the ratio of molecular--to--atomic gas  for the galaxy is typical of a
starburst. Thus in prograde grazing encounters, there appears to be
a time delay before conditions become suitable for a starburst. 
The length of the time delay depends on how rapidly post--encounter
evolution proceeds. This small
sample of three prograde encounters suggests that a central starburst may not
develop until near the end of the ocular phase.

It appears that the encounter caused a considerable
amount of gas to lose angular momentum and fall into
the center of NGC 5394, produced the lopsided distribution of gas  in
its inner disk, and may have increased the $M({\rm H_2)}/M({\rm HI)}$
ratio for the galaxy as a result of
the conversion of atomic to molecular gas in the high density central region.

\section{NGC 5395 Results}

We expected to find a compressed molecular ridge or clumped molecular gas 
at the following locations
in NGC 5395: (1) in the dominant spiral arm on the eastern side of NGC 5395,
since this coincides with a prominent radio continuum ridge 
(see Paper I); (2) in a stellar plus 
ionized gas ``shell--like'' structure ($20''$ in diameter, marked
on the NE side of NGC 5395 in Fig.~\ref{fig:cornell}) which Paper I 
interpreted as a caustic produced by temporary 
convergence of orbits; (3) in the prominent dust lanes along
the dominant spiral arm on the western side of NGC 5395. On the eastern side
of the galaxy, the dominant spiral arm is brighter in radio continuum and
\Halpha\ emission, whereas on the western side of the galaxy, it is
brighter in \HI. The face--on surface density of \HI\ plus helium (measured
with $11''$ resolution in Paper I) is 16 -- 20 \Msolpc2\ at Onsala position H 
(at the main dust lane 
on the western side of NGC 5395) and 10 \Msolpc2\ at the caustic.    
NGC 5395 has an inclination of 65\deg.

We compare the \12CO\ observations of NGC 5395 by the NRAO 12--m telescope 
(beam = $55''$ HPBW) \citep{ZHU99} and by Onsala
(beam = $33''$ HPBW) (Paper I) with the BIMA data
presented here. None of these data sets has complete coverage of the
galaxy. With three pointings at a spacing of $60''$, 
\citet{ZHU99} measured a total
integrated flux $S_{\rm CO}$ of $325 \pm 26$ Jy \kms, equivalent to 
$M(\rm H_2)$ = $7.1 \times 10^9$ \Msol\ with a standard value of $X$. They
then multiplied this flux by a factor of 1.7
to include CO emission outside of the region covered. Since $M$(HI) is
$1.4 \times 10^{10}$ \Msol\ (Paper I), applying this factor gives an
$M({\rm H_2})/M(\rm HI)$ ratio of 0.86, so NGC 5395, which does not have
a central starburst, is somewhat CO--rich compared to the typical spirals
in the survey by \citet{CAS98} but not as CO--rich 
as its companion. 

All three NRAO 12--m profiles have very broad CO linewidths \citep{ZHU99}.
Correcting these  for the \HI\ velocity gradient across the
CO beam and assuming that the CO emission fills the NRAO beamwidth
parallel to the kinematic major axis (position angle = 173\deg) results in 
a one-dimensional CO velocity dispersion of $\sim 80$ \kms. This is similar
to the \HI\ velocity dispersions of 50 -- 70 \kms\ found in Paper I.
As in the case of the \HI\ line widths, the large value for the velocity 
dispersion
probably results mainly because the line--of--sight intercepts gas at various
radial distances in the disk
and at various altitudes above the plane. Paper I presented evidence for 
three--dimensional disturbed structures in the \HI\ gas. The broad CO
linewidths suggest that molecular gas may also
participate in these three dimensional structures.

The Onsala and BIMA observations of NGC 5395 were not sufficiently sensitive to
detect the molecular gas with high velocity dispersion seen by the NRAO
12--m. The Onsala observations detected a CO component with low
velocity dispersion in the northern part of the ring/pseudo ring and
measured an $S_{\rm CO}$ of 126 Jy \kms\ from 
the two pointings (labelled H and I in Fig.~\ref{fig:cornell}) 
centered $40''$ north of the nucleus. At positions H and I, the Onsala line
profiles have
FWHM line widths of 77 \kms\ and 69 \kms, respectively, and central 
velocities of 3314 \kms\ and 3328 \kms\ (close to the \HI\ mean velocities).
From the \HI\ velocity field (Paper I), we find
a velocity gradient of 3.8 \kms\ arcsec$^{-1}$ at position H and
2.8 \kms\ arcsec$^{-1}$ at position I. If the CO emission fills the Onsala 
beam uniformly along the kinematic major axis, then the velocity gradient 
would produce a velocity 
difference across the beam of 125 \kms\ at position H
and 90 \kms\ at position I. Since these values are appreciably greater
than the FWHM linewidths observed by Onsala, the CO emission does not fill 
the Onsala beam uniformly along the gradient of the velocity field 
(approximately north--south). 
 The NRAO 12--m position
$60''$ north of the nucleus includes part of
Onsala positions H and I and the caustic feature. Its
line profile consists of a narrow component 
with $v$ = 3300 -- 3360 \kms\  (the same velocity range as the emission
found by Onsala at positions H and I) sitting atop the fainter,
broad component with FWHM = 230 \kms\ \citep{ZHU99}. 

\begin{figure}
\epsscale{0.5}
\plotone{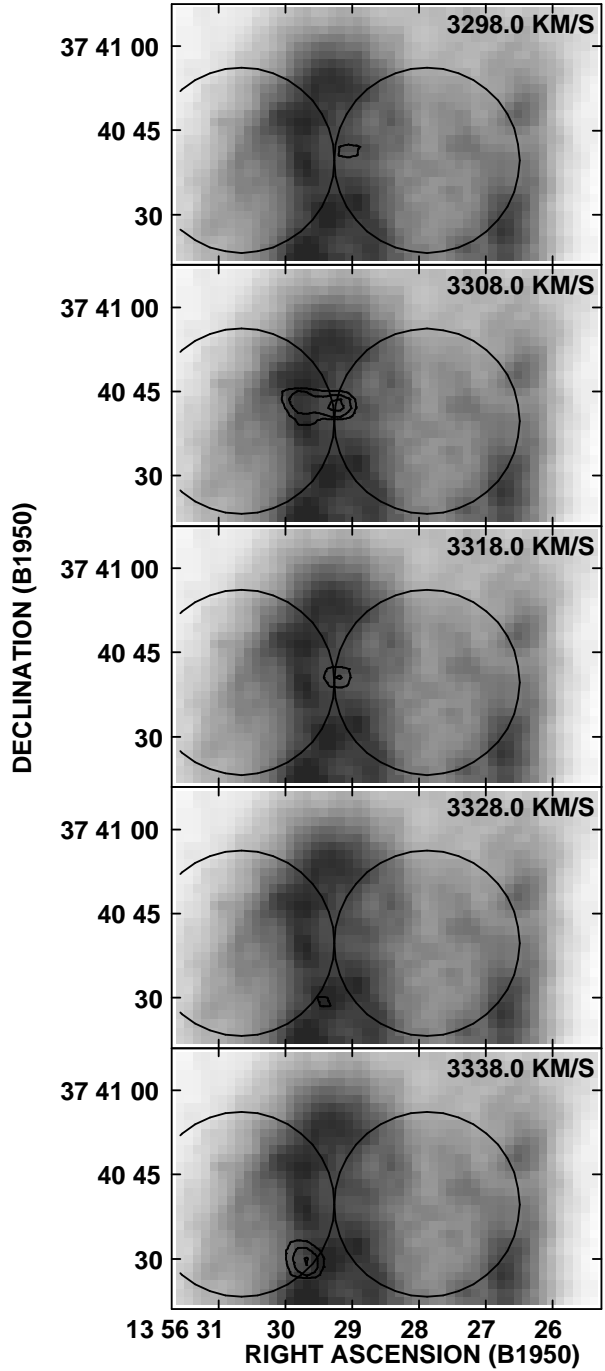}
\caption{A selection of channel maps from the BIMA CO cube overlaid
on a gray--scale display of the Digitized Sky Survey image of the northern part
of NGC 5395. The channel maps have been masked so that only detections or
possible detections are displayed. The contour levels are at 2, 3, and 4 
times the rms noise. The two large circles denote Onsala positions H and I,
labelled in Fig. 1.
\label{fig:N5395}
}
\end{figure}

The BIMA observations detect very little CO emission from NGC 5395.
In Figure~\ref{fig:N5395}, contours from  BIMA channel maps 
for the appropriate velocity range are overlaid on the Digitized Sky Survey 
image in gray--scale for a field containing the stellar caustic feature and 
positions H and I. 
This  displays 
the best detection of CO emission in the BIMA observations of
NGC 5395: a feature at the intersection
of positions H and I with maximum $I_{\rm CO}$ = 13 K \kms\ = 
$3.7 \times$ the rms noise and $S_{\rm CO}$ = 6 Jy \kms, which is
5\% of the $S_{\rm CO}$ measured by Onsala from the two pointings H and I.
 Its location mainly interior to the arc formed by the stellar caustic is
surprising as we had expected
to find the CO emission in this field along the stellar caustic. 
The only other possible CO
feature in Figure~\ref{fig:N5395} is, at best, a marginal detection on
the arm just south of the stellar caustic. 
The two channel at 
$2 \times$ rms noise upper limit in this field
corresponds to a face--on surface density, including helium, of 
15 \Msolpc2\ for a standard value of $X$.

    It is puzzling that BIMA missed almost all of the narrow velocity 
component to the CO emission at Onsala positions H and I.
If the emission measured by Onsala were clumped in a linear
feature $33'' \times 6.5''$, BIMA should have detected position H at
$8.9 \times$ the rms noise and position I at $4.7 \times$ the rms noise
(averaging the BIMA sensitivity over the Onsala aperture).   
The constraint imposed by
the Onsala linewidth implies that the CO emission detected by Onsala
has a N--S extent $\leq 20''$ at position H and $\leq 25''$ at position I.
If the CO feature has dimensions $20'' \times 33''$, then it could be 
below $2 \times$ the rms noise at position I in the BIMA observations
but it should have been present at $2.9 \times$ the rms noise at
position H in the BIMA data.  It may be that across the northern end of
the \HI, \Halpha, radio continuum ring/pseudo--ring of NGC 53945, there
is a $20''$ wide band of CO emission, possibly consisting of small, unresolved,
 molecular clouds, and that 
the E--W extent of the CO emission exceeds the largest
scale visible to BIMA in C configuration.

Since the BIMA observations detected very little CO emission from NGC 5395,
the molecular gas does not appear to be
strongly concentrated in compressed ridges where we expected to find 
strong shock fronts. We are left with the puzzle that the eastern half of
the ring/pseudo--ring is brighter in radio continuum and
\Halpha\ emission, whereas the western half has a higher column density of
total gas. 

\citet{ZHU99} found an $S_{\rm CO}$ of 
$166 \pm 17$ Jy \kms\ (380 \kms\ FWHM) from the central
NRAO 12--m pointing,\footnote{M. Zhu, private
communication, points out that the value of the peak
$T_{\rm mb}$ was misprinted in their Table 2A 
as 31.8 mK instead of 18.3 mK.} but neither Onsala nor BIMA detect CO
emission from the nuclear region of NGC 5395. The discrepancy between the
Onsala and NRAO 12--m results at this location can be explained if
most of the central CO emission seen by the NRAO 12--m 
comes from the dominant spiral arm on the eastern side of NGC 5395 
(not the nucleus) and fills the NRAO beamwidth parallel to the major axis 
of the galaxy.
The nucleus is also faint in the radio continuum (see Paper I).

\section{Conclusions}

The spiral galaxy NGC 5394 has suffered a recent, prograde, grazing encounter
with the spiral galaxy NGC 5395 (Paper I). In NGC 5394, we find 
(1) 80\% of the CO emission detected by BIMA comes from the central 800 pc
starburst region, (2) the gas distribution in the disk is lopsided,
with more CO, \Halpha, and \HI\ emission from the western or southwestern side,
 and (3) CO and \Halpha\ emission
are detected from the innermost western arm but not  
from the optically very bright, eastern or the outer--western inner--disk arms.
The \HI\ velocity dispersion in the disk is several times
higher than in undisturbed spiral galaxies (see Paper I), and the 
CO velocity dispersion at the innermost western arm is somewhat
higher than normal.
These properties appear to result from the tidal encounter: in an
encounter simulation that reproduces some of the main features of this galaxy 
pair, a considerable amount of gas in NGC 5394 falls into the central region 
early in the collision.
Previous studies (Paper I) concluded that the galaxies IC 2163, NGC 2535,
and NGC 5394 form three distinct stages in the development of structures in
prograde, grazing encounters, with NGC 5394 the most evolved interaction 
in this set. Of the three, only NGC 5394 has a central starburst with a
high concentration of molecular gas there, which suggests, for similar types
of encounters, a central starburst may not develop until near the end of the
ocular phase.

Star formation in the disk of NGC 5394 follows the gas distribution.
Upper limits on the CO integrated intensity at the eastern inner--disk arm
suggest that the instability parameter $Q_{\rm gas}$ is 
presently too high for significant star formation there. So the question of
why there is no evidence of ongoing star formation at the eastern
inner--disk arm is really the question of why the distribution of
molecular gas is lopsided. This is probably the result of an asymmetry in 
the collisional perturbation plus the gas losing angular momentum
in shocks and may be related to the west  versus east
time lag in the development of the ocular oval in the model.

\citet{KAU97} noted that widespread, high 
velocity dispersion in the \HI\ gas is 
observed in the disks of a number of interacting spiral galaxies during
an early phase of post--encounter evolution. With the BIMA observations of
NGC 5394, we have some evidence that enhanced turbulence in the disk
is also present in the 
molecular gas. The measured component of the velocity dispersion in NGC 5394
should represent mainly the $z$--component. In NGC 5395, the broad CO 
linewidths in the NRAO 12-m observations by \citet{ZHU99} 
indicate high velocity dispersions similar
to those in the \HI\ gas. Although molecular gas with high velocity dispersion
has been found previously in merger remnants and in the nuclear regions
of normal galaxies \citep{CAS91,BAL88}, our results indicate that 
molecular gas with high velocity dispersion may also occur in the disks
of galaxies involved in a grazing encounter.
 
If standard values are used for the conversion factor $X$ = 
$N({\rm H_2})/I_{\rm CO}$ and the dust--to--gas ratio, then the average
extinction of the NGC 5394 starburst implied by the integrated CO 
intensity $I_{\rm CO}$ is a factor of 3 -- 4 greater than that deduced from
comparison of the radio continuum, \Halpha, and 60 \micron\ luminosities.
From this we conclude that the value of $X$ in the starburst
is a factor of 3 -- 4 below the standard value. 
This adds the starburst region in NGC 5394 to the list of nuclear
starbursts in which the value of $X$ is
below the standard value.  
The CO position--velocity diagram of the nucleus of NGC 5394 reveals 
two separate velocity features very close
to the center. This may indicate a nuclear ring or the ``twin peaks'' of an
inner ILR or some destruction of CO at the nucleus by photodissociation
or consumption in star formation.

Very little of the CO emission detected from NGC 5395 in single--dish 
observations is seen in the BIMA data, and thus the molecular gas does
not appear to be strongly concentrated in compressed ridges. We are left with 
the puzzle that on the 
eastern side of NGC 5395, the ring/pseudo--ring is brighter in radio continuum
and \Halpha\ emission, whereas on the western side of NGC 5395, it has a higher
column density of total gas.

\acknowledgments

This research made use of the NASA/IPAC Extragalactic Database (NED), which is
operated by the Jet Propulsion Laboratory, California Institute of Technology,
under contract with the National Aeronautics and Space Administration. 
We thank Tyler Nordgren for providing the fits version of his $r$--band image.

\clearpage
\clearpage

\clearpage

\clearpage

\clearpage

\clearpage

\clearpage

\clearpage

\clearpage

\clearpage

\clearpage

\clearpage

\clearpage

\clearpage

\clearpage
\begin{deluxetable}{lcc}
\tablenum{1}
\tablewidth{0pt}
\renewcommand{\arraystretch}{.6}
\tablecaption{Properties of NGC 5394}
\tablehead{
\colhead{Parameter} &\colhead{Value}  &\colhead{Reference}}
\startdata
Morphological Type & SB(s)b pec & 1\\
Radio continuum nucleus\\
\hskip 24pt Right Ascension (1950.0) & $13^h 56^m 25\fs 10$ & 1\\
\hskip 24pt Declination (1950.0) & $37\deg\ 41' 45\farcs 7$ & 1\\
Isophotal major radius $R_{25}$ & $51''$ & 1\\
$B_{\rm T}$ & 13.70 & 1\\
Optical $v_{\rm sys}$ & $3451 \pm 12$ \kms\ & 1\\
Distance & 47 Mpc & 2\\
Linear scale & 230 pc arcsec$^{-1}$ & 2\\
Disk inclination & $15\deg \pm 2$\deg\ & 6\\
Projection line of nodes & $0\deg \pm 2$\deg\ & 2\\
$M(\rm {HI})$ & $7.3 \times 10^8$ \Msol\ & 2\\
Single--dish CO flux & 168 Jy \kms\ & 2\\
Semi--major axis of primary bar & $8''$ & 2\\
Position angle of bar major axis & 120\deg\ & 2\\
1.40 GHz flux density & 34.8 mJy & 2\\
4.86 GHz flux density & 15 mJy & 3\\
Central starburst\\
\hskip 24pt 1.45 GHz flux density & 29 mJy & 4\\
\hskip 24pt H$\alpha$ flux & $2.2 \times 10^{-13}$ erg cm$^{-2}$ s$^{-1}$ &
5\\
\enddata
\tablerefs{
(1) \citet{DEV91} and the NASA Extragalactic Database;
(2) \citet{KAU99}; (3) \citet{CON91};
(4) \citet{BEC95}; (5) \citet{KEE85}; (6) this paper.} 
\end{deluxetable}
\clearpage
\begin{deluxetable}{lc}
\tablenum{2}
\tablewidth{0pt}
\tablecaption{Summary of BIMA Observations}
\tablehead{
\colhead{Parameter} &\colhead{Value}}
\startdata
Date of Observation & 1999 June 4\\
Configuration & C\\
Synthesized Beam (HPBW, P.A.) & $6\farcs 6 \times 6\farcs 3, 86\deg$\\
Pointing Centers\\
\hskip 24pt R.A., DEC (1950) & $13^h\, 56^m\, 26\fs 152, +37\deg\ 41'\, 
27\farcs 14$\\
\hskip 24pt R.A., DEC (1950) & $13^h\, 56^m\, 27\fs 797, +37\deg\ 40'\,
47\farcs 21$\\
rms noise per channel  & 80 \mJybeam\ = 0.18 K\\
Equivalent $T_{\rm b}$ for 1 \Jybeam\ & 2.21 \K\\
Central velocity & 3497 \kms\ \\
Velocity resolution & 4.06 \kms\ \\
Hanning--smoothed channel width & 10 \kms\ \\
Total bandwidth & 908 \kms\ \\
Primary Beam (HPBW) & $100''$\\
Largest scale size visible to array & $40''$\\
\enddata
\end{deluxetable}
\clearpage
\begin{deluxetable}{lcc}
\tablenum{3}
\tablewidth{0pt}
\tablecaption{Radio, H$\alpha$, and CO Emission from Center of NGC 5394}
\tablehead{
\colhead{Wavelength} &\colhead{Source Size\tablenotemark{a}} 
&\colhead{Resolution}\\
& \colhead{(FWHM, P.A.)} &\colhead{(HPBW)}}
\startdata
$\lambda$ 20 cm radio continuum & $3\farcs 5 \times 3\farcs 4$, 66\deg\ & 
$5\farcs 4 \times 5\farcs 4$\\
H$\alpha$ & $3\farcs 5 \times 3\farcs 1$, 118\deg\ & 
$2\farcs 4 \times 2\farcs 4$\\
CO $J = 1\rightarrow 0$ & $4\farcs 7 \times 3\farcs 8$, 148\deg\ 
& $6\farcs 6 \times 6\farcs 3$\\
\enddata
\tablenotetext{a}{From Gaussian fit to emission, deconvolved from point spread
function} 
\end{deluxetable}

\end{document}